\documentclass[natbib,letterpaper,iop]{emulateapj}
\usepackage{amstext}
\usepackage{apjfonts}
\usepackage{amsmath}
\usepackage{amssymb}
\usepackage[colorlinks=true,linkcolor=blue,citecolor=blue]{hyperref}
\begin{document}

\newcommand{\beq}{\begin{equation}}
\newcommand{\eeq}{\end{equation}}
\newcommand{\beqar}{\begin{eqnarray}}
\newcommand{\eeqar}{\end{eqnarray}}
\newcommand{\Ra}{R_{\rm a}}
\newcommand{\rt}{r_{\rm t}}
\newcommand{\rp}{r_{\rm p}}
\newcommand{\Ma}{M_{\rm a}}
\newcommand{\Mbh}{M_{\rm bh}}
\newcommand{\vinf}{v_\infty}
\newcommand{\rhoinf}{\rho_\infty}
\newcommand{\mach}{\mathcal M_\infty}
\newcommand{\mhl}{\dot M_{\rm HL}}
\newcommand{\lhl}{\dot L_{\rm HL}}
\newcommand{\Hrho}{H_\rho}
\newcommand{\Rs}{R_{\rm s}}
\newcommand{\erho}{\epsilon_\rho}
\newcommand{\cs}{c_{\rm s,\infty}}
\newcommand{\ehl}{\dot E_{\rm HL}}

\title{The Close Stellar Companions to Intermediate Mass Black Holes} 

\author{Morgan MacLeod$^{1}$, Michele Trenti$^{2}$, and Enrico Ramirez-Ruiz$^{1}$}
\altaffiltext{1}{Department of Astronomy \& Astrophysics, University of California, Santa Cruz, CA 95064}
\altaffiltext{2}{School of Physics, The University of Melbourne, VIC 3010, Australia}
   
\begin{abstract} 
When embedded in dense cluster cores, intermediate mass black holes (IMBHs) acquire close stellar or stellar-remnant companions. These companions are not only gravitationally bound, they tend to hierarchically isolate from other cluster stars through series of multibody encounters. In this paper we study the demographics of IMBH companions in compact star clusters through direct $N$-body simulation. We study clusters initially composed of $10^5$ or $2\times 10^5$ stars with IMBHs of 75 and 150 solar masses, and follow their evolution for 6-10 Gyr. A tight innermost binary pair of IMBH and stellar object rapidly forms. The IMBH has a companion with orbital semi-major axis at least three times tighter than the second-most bound object over 90\% of the time. These companionships have typical periods of order years and are subject to cycles of exchange and destruction. The most frequently observed, long-lived pairings persist for $\sim 10^7$ yr.  The demographics of IMBH companions in clusters are diverse; they include both main sequence, giant stars, and stellar remnants. 
Companion objects may reveal the presence of an IMBH in a cluster in one of several ways. 
Most-bound companion stars routinely suffer grazing tidal interactions with the IMBH, offering a dynamical mechanism to produce repeated flaring episodes like those seen in the IMBH candidate HLX-1.
Stellar winds of companion stars provide a minimum quiescent accretion rate for IMBHs, with implications for radio searches for IMBH accretion in globular clusters. 
Finally, gravitational wave inspirals of compact objects occur with promising frequency. 
\end{abstract}


\maketitle

%
% INTRO
%
\section{Introduction}\label{sec:intro}
Globular clusters (GCs),  dense young star clusters, and the compact  nuclear star clusters of low-mass galaxies are environments of extreme stellar density. Their typical sizes of parsecs and masses $\gtrsim 10^5 M_\odot$ imply stellar number densities  between $10^3$ and $10^6$ times that of the solar neighborhood  and velocity dispersions of tens of km s$^{-1}$ \citep{2003gmbp.book.....H,2012AdAst2012E..15N}.   With similar ratios of stellar mass to system mass, stellar escape velocity to velocity dispersion, and stellar lifetime to system relaxation time, these dense stellar systems have broad dynamical similarities despite their disparate environments. These dense clusters are environments of  intense stellar interactivity, where single and binary stellar evolution and gravitational dynamics intertwine to shape the long-term evolution of the cluster as a whole \citep{2003gmbp.book.....H,2003gnbs.book.....A,2010RSPTA.368..755K}. 

There has long been speculation that these dense stellar systems might harbor intermediate mass black holes (IMBHs) with masses between that of the known populations of stellar mass and supermassive black holes (BHs) \citep[e.g.][]{1978RvMP...50..437L,2002MNRAS.330..232C}. If the $M_{\rm BH}-\sigma$ relation of  supermassive BH mass and velocity dispersion were to extrapolate to objects as small as GCs and nuclear clusters, it would imply that these stellar systems should host BHs in the IMBH mass range \citep[e.g.][]{2000ApJ...539L...9F,2013A&A...555A..26L,2015ApJ...809L..14B}.
In young clusters of low metallicity, evolving massive stars might preferentially leave behind massive remnant BHs, with masses $\gtrsim 10^2 M_\odot$ \citep{2013MNRAS.429.2298M,2015MNRAS.451.4086S}. Through dynamical interaction these BHs could acquire and accrete mass from companion objects \citep{2013MNRAS.429.2298M,2014ApJ...794....7M,2014MNRAS.441.3703Z} implying a self-consistent channel by which the BH could grow into the IMBH mass range as the cluster evolves.

There is growing evidence that some nuclear star clusters in low-mass galaxies host BHs in the IMBH mass range \citep[e.g.][]{2012AdAst2012E..15N}.  These sources have been discovered primarily in the fraction of systems where the central BH is active. For example, \citet{2008ApJ...678..116S} study the coincidence of BH activity and nuclear star clusters, and find that a large fraction of all nuclear star clusters show signs of BH activity ($>50$\% of those with masses $\gtrsim10^7 M_\odot$). \citet{2010ApJ...714..713S} found a BH mass of $\sim 5\times 10^5 M_\odot$ in the NGC 404 nucleus.  
By studying the active BH fraction in a large sample of Sloan Digital Sky Survey dwarf galaxies, \citet{2013ApJ...775..116R} find a $\sim 0.5$\% active fraction in their sample of $10^{8.5}-10^{9.5} M_\odot$ dwarfs with median BH masses of $\sim 2 \times 10^5 M_\odot$. \citet{2013ApJ...775..116R} note that sensitivity limits their ability to detect active BHs of less than $10^5 M_\odot$, even if they are emitting near their Eddington luminosities. Even so,  their data suggest a similar active fraction among these low mass BHs to that of $10^7 M_\odot$ BHs.    More recently, \citet{2015ApJ...809L..14B} report a $5\times 10^4 M_\odot$ IMBH in the dwarf galaxy RGG 118 accreting at 1\% its Eddington luminosity. The fact that this BH lies along the extrapolation of the $M_{\rm BH}-\sigma$ relation to low masses suggests that BHs in this mass range may be common in dwarf galaxies, with their detection hampered only by their low characteristic luminosities.

The evidence for the presence or absence of IMBHs in GCs remains contested. An IMBH, unless illuminated through accretion, would be be dark \citep[e.g.][]{2012ApJ...750L..27S,2009ApJ...697L..77R,2011ApJ...730..145V}. This leaves only the IMBH's gravitational interaction with surrounding stars as a tracer of its presence. Searches for peaks in the central velocity dispersion or surface brightness of GCs due to orbital motion around an unseen IMBH have returned contested results \citep[e.g.][]{2005ApJ...634.1093G,2007ApJ...668L.139L,2008ApJ...676.1008N,2010A&A...514A..52M,2010ApJ...719L..60N,2011ApJ...743...52N,2011A&A...533A..36L,2012A&A...538A..19J,2012A&A...542A.129L,2012A&A...543A..82L,2013A&A...552A..49L,2013A&A...554A..63F,2014A&A...566A..58K,2015A&A...581A...1L}, with a fundamental challenge  being that unless the IMBH is a large fraction (e.g. $\gtrsim 1$\%) of the cluster mass very few stars will trace orbits dominated by the IMBH's gravity \citep[e.g.][]{2010ApJ...720L.179V,2013ApJ...768...26U}. This challenge has lead to theoretical and observational efforts to identify other dynamical tracers of the presence of IMBHs in GC systems \citep[e.g.][]{2005ApJ...620..238B,2007PASJ...59L..11H,2007MNRAS.379...93H,2008ApJ...686..303G,2009ApJ...699.1511P,2010MSAIS..14...59P,2010ApJ...710.1063V,2010ApJ...713..194B,2012ApJ...751L..12S,2014MNRAS.444...29L}. 

In this paper, we consider a stellar dynamical implication of the possible presence of IMBHs in dense cluster cores: stars bound tightly to the IMBH.   IMBHs in clusters acquire a number of gravitationally bound stars that make up their sphere of influence \citep{1972ApJ...178..371P,1976ApJ...209..214B,2004ApJ...613L.109P,2004ApJ...613.1133B,2004ApJ...613.1143B}. These stars orbit under the combined influence the IMBH and all of the other bound stars \citep{1976MNRAS.176..633F}. The stellar dynamics of the relatively small number of stars in the sphere of influence region is complex, and relies on the interface of stars' orbital, dynamical relaxation, and evolution timescales. In simulation, the most accurate (but also computationally expensive) approach to treating the multiphysics sphere of influence dynamics is direct $N$-body evolution of the stars' orbital motion \citep{2003gnbs.book.....A}. 
Companion objects to IMBHs have been mentioned in previous studies  \citep[e.g.][]{2004ApJ...613.1143B,2006ApJ...642..427B,2006MNRAS.372..467B,2013A&A...557A.135K,2014MNRAS.444...29L,2015MNRAS.454.3150G}, however, in this paper we present the first systematic study of IMBH companions in a dense cluster context that takes advantage of the accuracy of direct $N$-body integration.

Stellar objects or stellar remnants tightly bound to an IMBH are of extreme astrophysical interest because of their ability to reveal the presence of the otherwise dark BH. The category of ultraluminous X-ray (ULX) sources is particularly interesting because their X-ray luminosities of $\gtrsim 10^{39}$~erg~s$^{-1}$ preclude an Eddington limited stellar mass BH source \citep[e.g.][]{2003MNRAS.344..134H,2014Natur.513...74P}. 
In dense clusters, stars can be scattered into close, disruptive encounters with IMBHs, powering luminous but brief flares of ULX luminosity \citep{2009ApJ...697L..77R}. 
If, instead, stars are captured into tight orbits and induced to Roche lobe overflow, persistent ULXs could conceivably be the very high mass analogs X-ray binaries \citep{2004ApJ...604L.101H,2004ApJ...616L.119L,2004MNRAS.355..413P,2006ApJ...642..427B,2006MNRAS.370L...6P,2006MNRAS.372..467B}. 
If less tightly bound to the IMBH, a stellar companion could still fuel accretion through its stellar wind \citep{2005MNRAS.360L..55C,2006MNRAS.366..358C,2008MNRAS.383..458C,2014ApJ...788..116M}. 

The source HLX-1 \citep{2009Natur.460...73F} is particularly dramatic, because its lightcurve shows repeated flaring episodes during which the source undergoes state transitions, strongly suggesting an  IMBH as the central engine \citep{2009ApJ...705L.109G,2011AN....332..392F,2011ApJ...734..111D,2011ApJ...743....6S,2012Sci...337..554W,2014ApJ...780L...9W,2014MNRAS.437.1208F}.  The roughly periodic flaring behavior has also lead to an association of the behavior with a IMBH plus star system in an eccentric orbit -- where close pericenter passages might account for the brightening \citep{2011ApJ...735...89L,2013MNRAS.428.1944S,2014ApJ...788..116M,2014ApJ...793..105G}. The association of the object with a compact stellar system strengthens the interpretation that the accretion flares might be powered by a stellar companion \citep{2010ApJ...721L.102W,2012ApJ...747L..13F,2012MNRAS.423.1309M,2013MNRAS.433..849M,2013A&A...559A.124M,2013ApJ...768L..22S}.  An avenue for these repeated, grazing tidal interactions with supermassive BH hosts has been suggested by \citet{2013ApJ...777..133M}, where stars evolve to `spoon-feed' portions of their envelope to the supermassive BH over many passages. In the galactic center context, \citet{2014ApJ...786L..12G} have proposed this channel as a dynamical mechanism to produce the tightly-bound G2 gas cloud. However, the dynamical processes that lead an IMBH, like the putative BH in HLX-1, to acquire then strip mass from a tightly bound companion remain poorly understood. 

This paper studies the dynamical processes by which IMBHs in dense clusters acquire, retain, and lose close companion stars. We find that a stellar object among the sphere of influence members tends to segregate to substantially tighter separations than the other bound stars. These companion objects form a hierarchically separated binary with the IMBH, and persist until they are replaced in an exchange interaction or destroyed through direct interaction with the IMBH. We characterize the orbital demographics and statistics of these companion objects using the $N$-body simulations presented in this paper. In Section \ref{sec:method} we describe our direct $N$-body simulation method and parameter choices. In Section \ref{sec:comp} we explore the stellar dynamics, populations, and processes governing tightly-segregated companions to the IMBH. In Section \ref{sec:diff} we examine the dependence of our simulated populations of companion on some key cluster parameters.  In Section \ref{sec:disc} we discuss some astrophysical implications of our results, particularly the role that these tightly bound companions might play in revealing the presence of otherwise hidden IMBHs. Finally, in Section \ref{sec:conclusion} we conclude.

%
% METHOD
%
\section{Direct $N$-body Simulations of Clusters Hosting IMBHs}\label{sec:method}

\subsection{Simulation Method}

In this work we use the direct-summation $N$-body star cluster dynamics code NBODY6 \citep{1999PASP..111.1333A,2003gnbs.book.....A} in a version enabled for GPU-accelerated calculations \citep{2012MNRAS.424..545N}\footnote{Version: August 1, 2014.} to follow the evolution of star clusters in a realistic tidal field following the implementation of \citet{2006MNRAS.368..677H} and subsequent papers \citep{2007MNRAS.374..344T,2007MNRAS.374..857T,2008IAUS..246..256T,2008MNRAS.387..815T,2008ApJ...686..303G,2009ApJ...699.1511P,2010ApJ...708.1598T,2010ApJ...720L.179V,2013MNRAS.435.3272T}.  
Simulations containing $N$ stellar particles are performed in $N$-body units where $G = M_{\rm T}=-4E_{\rm T}=1$, in which $M_{\rm T}$ and $E_{\rm T}$ are the total mass and energy of the cluster initial conditions \citep{1986LNP...267..233H}. The corresponding dynamical time unit is 
\beq\label{tdyn}
t_{\rm d} = G M_{\rm T}^{5/2} / (-4 E_{\rm T})^{3/2}.
\eeq
The stellar cluster's dynamical state relaxes over approximately a relaxation time,
\beq
t_{\rm rh} = \frac{0.138 N r_{\rm h}^{3/2} } {\ln(0.11 N)},
\eeq
 defined using the half mass radius, $r_{\rm h}$ \citep{1987degc.book.....S,2010ApJ...708.1598T}.
The external tidal field of the host galaxy is implemented following \citet{2007MNRAS.374..344T} and the reader is directed to section 2 of that paper for details.

The NBODY6 distribution embeds the SSE and BSE codes of \citet{2000MNRAS.315..543H,2002MNRAS.329..897H} as a simplified, on-the-fly implementation of stellar evolution\footnote{Compared to the more recent stellar evolution calculations of \citet{2015MNRAS.451.4086S}, the mass distribution of stellar remnants in our approach is favors lower values, especially at low metallicities, with a potential impact on the dynamics being the number of objects with mass much greater than the average stellar mass (e.g., see \citealt{2014ApJ...794....7M}).}. The binaries in our clusters evolve with the standard BSE implementations of binary evolution. NBODY6 simulations include default binary evolution processes implemented in BSE including tidal circularization, magnetic braking, Roche Lobe overflow, and binary coalescence \citep{2002MNRAS.329..897H}.

We initialize our models with an IMBH of mass $\Mbh$ at rest at the system center of mass. This IMBH particle then evolves identically to the remaining particles. In this sense, our simulations do not self-consistently account for the formation of such an IMBH, instead we aim to simulate its subsequent dynamical evolution in the cluster context. Because the IMBH is more massive than any of the stellar particles it sinks toward the cluster center, motivating our initial condition. We find within the first core relaxation time the BH acquires non-zero energy relative to the cluster center of mass through strong interactions with cluster core stars -- rapidly erasing any imprint of this initial condition.  We include gravitational wave energy and angular momentum losses from any very compact binaries following the \citet{1964PhRv..136.1224P} formulation, which is particularly relevant for the IMBH and its companion \citep[e.g.][]{2002MNRAS.330..232C,2004ApJ...616..221G,2006ApJ...640..156G,2014MNRAS.444...29L}.

Close interactions with the IMBH can lead to the tidal destruction of a star or the inspiral and merger of a compact object. We handle those events as follows in our simulations. For stars, we delete any star passing within one tidal radius,
\beq\label{rt}
\rt = (\Mbh/M_*)^{1/3} R_*,
\eeq
 of the BH at pericenter.  
In a tidal disruption event, the star is often on a weakly-bound or near-parabolic orbit \citep{1976MNRAS.176..633F,1977ApJ...211..244L}. In this case approximately half of the tidal debris remains bound to the BH, while the other half is ejected \citep{1988Natur.333..523R}. We model this behavior by adding half of the disrupted star's mass to the IMBH's mass following a disruption event. 
An ideal implementation of the stellar tidal disruption process would account for the stellar structure in determining the degree of mass loss, and would also allow for partial tidal disruptions \citep{2013ApJ...767...25G}. Partial tidal stripping is particularly relevant to evolved stars which contain a dense, tidally invulnerable core \citep{2012ApJ...757..134M,2013ApJ...777..133M,2014ApJ...788...99B}.  However, in this work we adopt the simplified, full-disruption model described above. 
We implement gravitational wave inspiral and merger events through the inclusion of gravitational wave corrections to the two-body equation of motion of tight pairs of compact objects \citep{1964PhRv..136.1224P} . Rather than allowing merging particles to spiral to arbitrarily small separations, we instead merge pairs of compact objects whose inspiral time becomes shorter than 1 $N$-body time unit, equation \eqref{tdyn}. We do not apply a kick to the remnant $\Mbh$, although such a kick may be relevant for some mergers \citep{2008ApJ...686..829H,2013A&A...557A.135K}.

We have implemented several diagnostics to allow us to better study close companions to the IMBH. We record all three-body encounters in which a third object approaches with pericenter distance within three times the semi-major axis of the most-bound companion to the IMBH. We also include the parameters of all tidal disruption and merger events described above. For close tidal passages that do not lead to disruption, those with pericenter distances $\rt<\rp<10^3 \rt$ we also record the system parameters.  While we record full-cluster mass, position, and velocity snapshots every 10 $N$-body time units, we record the IMBH's position and velocity every 0.1$N$-body time unit. Every $N$-body time unit we record the parameters of the most tightly bound partner to the IMBH. 

\subsection{Numerical Simulations}

\begin{table}[tbp]
\begin{center}
\begin{tabular}{cccccccc}
\tableline\tableline
Name & Number & $N$ & $W_0$ & $r_{\rm h,0}$ & $M_{\rm bh,0}$ & $t_{\rm max}$ & $\sigma_{\rm k}/\sigma_*$ \\
 & & & & [pc] & $[M_\odot]$ & [Gyr] & \\
\tableline
A & 3 & 100k & 7 & 2.3 & 150 & 6 & 2.5  \\
B & 3 & 100k & 7 & 2.3 & 150 & 6 & 1  \\
C & 3 & 100k & 7 & 2.3 & 75 & 9 & 2.5  \\
D & 4 & 200k & 7 & 2.3 & 150 & 10.4 & 2.5  \\
\tableline
\end{tabular}
\caption{Table of  $N$-body simulation groups A-D, the number of simulations run per group, as well as their parameters and initial conditions. }\label{table:sims}
\end{center}
\end{table}

In this paper, we present 4 groups of simulations with differing initial parameters. Simulations within each group have statistically different realizations of the initial conditions. Table \ref{table:sims} defines the properties these simulation groups. 
We use the McLuster  code \citep{2011MNRAS.417.2300K} to generate the initial stellar distribution following \citet{1966AJ.....71...64K} models with $W_0 = 7$. 
Our simulations include a IMBH which is initialized at the center of mass with zero velocity and $N=1-2 \times 10^5$ stars.  The size of our clusters is constrained by the computational expense of the direct $N$-body method.  These clusters are thus at the low mass end of the spectrum of GC masses, and a factor of 10-100 less massive than typical nuclear star clusters.  In later sections we consider how our results might scale to larger $N$ clusters.

The stars in our initial conditions follow a \citet{2001MNRAS.322..231K} initial mass function (IMF), within the mass range $0.1 M_\odot - 30 M_\odot$, with no binaries. The metallicity is one-tenth solar.  We initialize our clusters as tidally-underfilling by a factor of 1.5. In $N$-body units, the initial tidal radius is 10.455 rather than 6.95 for $W_0=7$ in our simulations \citep[compare to Table 1 of ][]{2007MNRAS.374..344T}. This allows for some cluster expansion due to mass loss before stars begin to escape. We consider cases of different IMBH mass, varied between 0.13\% and 0.26\% of the initial cluster mass.  

Finally, we consider two different cases for the velocity kicks imparted to stellar remnant NSs and BHs \citep[for a recent discussion, see][]{2015MNRAS.449L.100C}.  Rather than choosing absolute kick velocities based on observational constraints, here we choose velocities relative to the cluster velocity dispersion in order to produce different retention fractions of stellar remnants. This strategy is needed because the clusters we are able to simulate with the direct $N$-body method are small relative to realistic dense star clusters. In both cases remnants are given a kick drawn from a Maxwellian distribution with sigma of either 1 or 2.5 times the initial cluster velocity dispersion, $\sigma_* = \sqrt{G M_{\rm T} / r_{\rm h} }$. These kick distributions imply that 19.8\% or 1.6\%, of kicked objects have kick velocities less than $\sigma_*$, which in the absence of binaries, is proportional to the fraction of retained remnants. 

\subsection{Methodological Comparison to Previous Work}

Some previous work has mentioned or studied close companion stars to IMBHs in GCs. Here we describe the methodological differences in our new simulations as they compare to this work.  \citet{2004ApJ...613.1143B} run direct $N$-body simulations of clusters containing IMBHs and up to 131K stars. They plot several sequences of semi-major axis of tightly bound stars as a function of time from their simulations containing IMBHs (their Figures 14 and 15). These sequences show a long-lived most-bound companion in several cases with semi-major axis considerably smaller than the other stars and hint at the possible universality of IMBH companions. 
\citet{2006MNRAS.372..467B} focus on the production of mass-transferring  binary pairs containing the IMBH and a star. They run $N$-body simulations including approximate tidal dissipation and gravitational wave radiation from stellar orbits and study the statistics of tidally-captured stars in young (<12 Myr) clusters as possible ULX progenitors. 
\citet{2013A&A...557A.135K} also model small, young (10 Myr age) clusters initially containing $N$=32k stars using  direct $N$-body integration. The clusters  a massive IMBH with $M_{\rm bh} = 500 - 1000 M_\odot$. They study one simulation in detail, in which the IMBH captures a stellar BH companion, goes through several cycles of exchange, and eventually inspirals and merges. \citet{2013A&A...557A.135K} argue that the merger generates sufficient gravitational wave recoil to eject the IMBH from their low-mass simulation cluster.  
 \citet{2013MNRAS.429.2298M}, \citet{2014ApJ...794....7M}, and \citet{2014MNRAS.441.3703Z} study the dynamics of massive stellar-mass BHs (up to $10-100 M_\odot$) in a large sample of $N$-body calculations of young star clusters with a range of metallicities and $N=5500$ stars.  

\citet{2014MNRAS.444...29L} also run $N$-body simulations containing IMBHs with between $N$=32k and $N$=131k stars, but they do not include gravitational radiation from compact binaries. After the formation of an IMBH-BH compact binary,  the eccentricity of this innermost binary varies over time. The binary passes many times through orbital configurations that should have driven it to merger. As a result, \citet{2014MNRAS.444...29L}'s simulations show artificially long-lived companion objects, and a single companion persisting throughout the cluster evolution. 
Recently, \citet{2015MNRAS.454.3150G} have used the MOCCA Monte Carlo Cluster simulation code to study accretion-fed IMBH growth over GC lifetimes. Their simulations, particularly their ``slow" scenario with a single massive BH, show the cycles of replacement of the innermost object missing from earlier $N$-body work. However, as \citet{2015MNRAS.454.3150G} describe, the Monte Carlo code is not well-suited to studying the detailed dynamics small number of stars tightly bound to the IMBH, nor is it well-suited to the inclusion of a single, massive object. 

\citet{2006ApJ...642..427B} adopt a slightly different approach. Rather than simulate an entire GC, they use a binary evolution code to evolve the IMBH and any companions with a fixed background of scatterers drawn from a GC core distribution function.   While this approach facilitates modeling a larger number of IMBH pairs and a detailed treatment of IMBH-star interactions, it does not capture the dynamical feedback of the IMBH and its partner on the host cluster.

The simulations presented in this paper were designed specifically to study close companion objects to the IMBH in realistic GCs. This involved combining some of the best aspects of the previous work described above.  In particular, we take advantage of the accurate dynamics of the $N$-body methods, and implement specific interaction outcomes for close interactions between the IMBH and stellar objects like tidal disruptions and gravitational wave inspirals. We model an IMBH which is a realistically  small fraction (< 0.3\%) of the total cluster mass. This approach limits our study to relatively small GCs with $N_0 \sim 10^5$, with their proportionately low-mass IMBHs, $\Mbh \sim 10^2 M_\odot$, but allows us to follow their long-term evolution over 6-10 Gyr without approximation. 
 In Section \ref{sec:diff}, we discuss some considerations in scaling the lessons learned from our simulations to more massive clusters or IMBHs.

%
% Cluster Evol. 
%
\subsection{Cluster Global Evolution}

Our simulated clusters exhibit global evolution similar to those of others studied in detail combining IMBHs and a tidal field \citep[e.g.][]{2007MNRAS.374..857T,2008ApJ...686..303G,2010ApJ...708.1598T,2013A&A...558A.117L,2015MNRAS.454.3150G}. Thus, we comment only briefly on the generic properties of our model clusters to set the stage for our consideration of the dynamics of their IMBH-hosting cores. Clusters in a tidal field slowly dissolve as (typically low-mass) stars escape from the system.  Our models of simulation groups A and B are run for 6 Gyr, during which time the total number of stars decreases to $\approx 2 \times 10^4$. Simulation groups C and D are run to the point of dissolution in the tidal field, which occurs after 9 and 10.4 Gyr, respectively.  Mass loss from stellar evolution rapidly decreases the cluster mass by $\sim30$\% in the first 50 Myr, then proceeds increasingly gradually as time goes on.  The combination of these effects dictate that the mass range of typical cluster stars narrows over the course of the GC's evolution. At early times, stars are the most-massive objects other than the IMBH, while at later times, for example $\gtrsim 1$ Gyr, the turnoff mass has decreased to $<3M_\odot$, and stellar remnants are the most massive objects.   \citet{2013A&A...558A.117L}'s Figure 1 gives an excellent overview of the time-evolution of $N$-body models of GCs containing IMBHs.

\begin{figure*}[tbp]
\begin{center}
\includegraphics[width=0.99\textwidth]{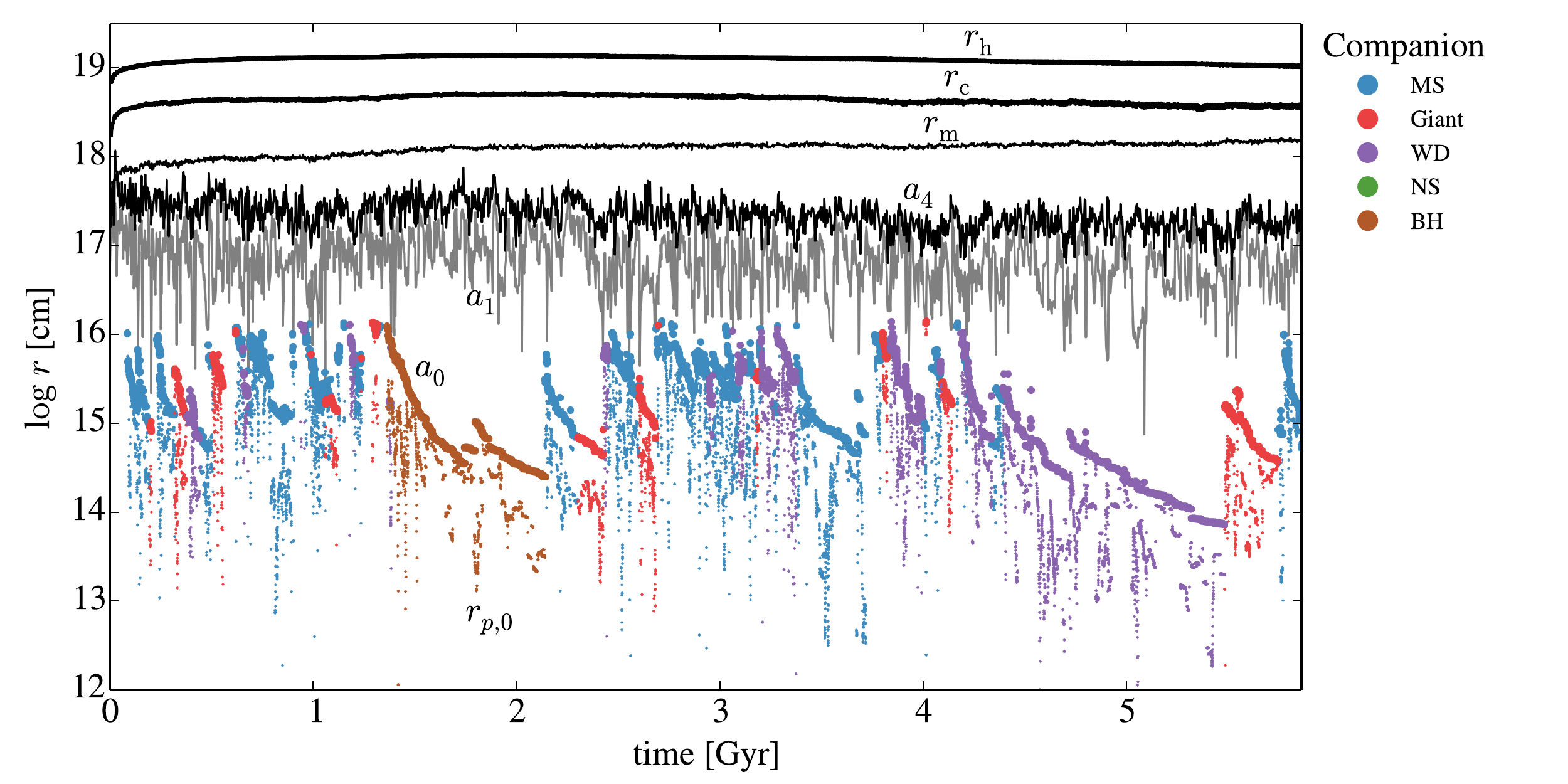}
\caption{Characteristic radii from the IMBH in an example cluster evolution (from simulation group A).  This diagram shows the evolution of the half-mass, $r_{\rm h}$, and core, $r_{\rm c}$, radii, along with the distance that encloses a IMBH-mass of stars, $r_{\rm m}$. Interior to these radii, we plot the semi-major axes of the 5th most-bound object, $a_4$, the second most bound, $a_1$, and the most bound, $a_0$. We plot the pericenter distance of the most-bound object, $r_{\rm p,0}$, as well. Coloring of the most-bound object denotes its stellar-evolutionary type, labeled in the key on the right-hand side. Stellar types include main sequence (MS) and giant stars. The remnant types include white dwarfs (WD), neutron stars (NS) and black holes (BH).  
}
\label{fig:radii_ks}
\end{center}
\end{figure*}

The IMBH sinks nearly to the cluster center of mass, as for example, can be seen in Figure 2 of \citet{2013A&A...557A.135K}. Once in the core, dynamical interactions with the IMBH and its companion object act as a heat source for the cluster, stalling core collapse or causing the system to expand \citep[e.g.][]{1996NewA....1...35Q,2003ApJ...599.1129Y,2007MNRAS.374..857T,2013A&A...558A.117L}. Figure \ref{fig:radii_ks} shows an example cluster evolution from simulation group A. The half-mass and core radii expand initially in response to stellar evolution mass loss and mass loss from kicked supernova remnant objects. These characteristic radii then stabilize for the remainder of the simulation and the cluster neither expands nor contracts as energy loss balances dynamical heating.  Over time, the IMBH grows by accreting the debris of tidally disrupted stars and by swallowing inspiralling compact objects. The individual sequences are, of course, variable, but the IMBHs in our clusters grow by $\sim 30$\% over the evolutionary span of the cluster lifetime.

%
% BH Companions: Case A
%
\section{Close Companions to the IMBH}\label{sec:comp}

In this section, we examine the properties of close companions to a cluster IMBH using our fiducial simulation group A described in the previous section and in Table \ref{table:sims}.  The qualitative behavior of each of the simulation groups is similar, and we defer to Section \ref{sec:diff} to  consider differences in our results between the cluster parameters represented in simulation groups B-D. 

\subsection{Stars Bound to the IMBH}
The process of capture of a close companion begins with the two body relaxation of the cluster. 
The stellar cluster's dynamical state relaxes over approximately a relaxation time, $t_{\rm rh}$  \citep{1987degc.book.....S}.
This relaxation drives the stellar distribution surrounding the IMBH to higher and higher densities, until a steady-state distribution is reached \citep{1972ApJ...178..371P,1976ApJ...209..214B,1976MNRAS.176..633F,1977ApJ...211..244L}. 
Because the IMBH mass is much greater than the average stellar mass, many stars can be gravitationally bound to the IMBH at a given time. For these stars,
\beq
\varepsilon = -{G \Mbh \over r_{\rm bh}} + {1 \over 2} v_{\rm bh}^2 < 0, 
\eeq
where we've written their orbital energy per unit mass $\varepsilon $ with respect to IMBH in terms of their instantaneous separation $r_{\rm bh}$ and velocity $v_{\rm bh}$ relative to the IMBH. 
As such, these are the stars that occupy a phase space where the IMBH is energetically dominant in their orbits. In general, these stars satisfy, 
\beq\label{rh}
a \lesssim r_{\rm h} = { G \Mbh / \sigma_{\rm c}^2 },
\eeq
where $a = G \Mbh / 2 \varepsilon$ is the semi-major axis of the bound star, and $\sigma_{\rm c}$ is the core velocity dispersion. 
While many stars may exist in such bound orbits, we would not describe each of these as a companion to the IMBH. 

Instead,  we will consider the IMBH to have a companion when its most-bound star is significantly more tightly bound than the second most bound. One way to express this is by comparison of those star's semi-major axes (or equivalently orbital energies) with respect to the IMBH. The condition can be written $a_0 / a_1 \ll 1$ which implies $ \varepsilon_0  /  \varepsilon_1 \gg 1$. Here the coefficient 0 denotes the most bound and coefficient $i$ denotes the $(i+1)$-th most tightly bound object.  In these cases, the most-bound object has hierarchically separated from the rest of the cluster and sphere of influence stars. The remaining stars then orbit in the combined potential of the IMBH and its companion. The hierarchically isolated companion can exist relatively unperturbed until it suffers a strong encounter. 

\subsection{Companion Capture \& Orbital Hardening}

How does the most-bound star end up in a orbit hierarchically isolated from the other cluster stars? 
We explore this question visually in Figure \ref{fig:radii_ks}. This diagram shows characteristic radii during a timeseries of one of the case A simulations.  From top to bottom, the radii plotted are the cluster half-mass radius, $r_{\rm h}$, the core radius, $r_{\rm c}$, the radius enclosing a IMBH-mass of stars $r_{\rm m}$. Next we show the characteristic radii of some of the most-bound companions to the IMBH. These include the semi-major axis of the 5th-most bound star $a_4$, the 2nd-most bound star, $a_1$, and the most bound, $a_0$. We also show the pericenter distance of the most-bound object, $r_{\rm p,0}$. We color the most-bound object depending on its stellar-evolutionary type. 

The larger radii, those that describe the bulk of the cluster, like the core and the half-mass radii are nearly constant in time.  As the radii get smaller, and describe fewer and fewer particles, a higher degree of variability is seen.  The semi-major axis of the most-bound object however, exhibits qualitatively different behavior than the other radii, or even the other tightly bound objects (as shown by $a_1$ and $a_4$). The most-bound star goes through a cyclic behavior of tightening semi-major axis followed by replacement with a new object. This cycle is seen to repeat throughout the timeseries of Figure \ref{fig:radii_ks}. 

The objects energetically bound to the IMBH all represent pairings that are `hard' relative to the cluster background energy \citep{1975MNRAS.173..729H,1983ApJ...268..319H}. In clusters, binaries that are hard tend to get harder through interactions with remaining cluster stars \citep{1975MNRAS.173..729H}. This process can be understood as a statistical trend toward energy equipartition between the binary and the background as kinetic energy tends to transfer from the binary components to the perturber, ejecting it faster than it entered \citep{1975AJ.....80..809H,1975AJ.....80.1075H}. In this picture, each encounter leaves the IMBH plus companion pair (on average) somewhat more tightly bound, and ejects the perturbing object with high velocity \citep{1974ApJ...190..253S}.

Among the multiple stars bound to the IMBH, only the most-tightly object is `hard' relative to the other bound stars. The orbits of less-bound objects cannot easily perturb the innermost orbit. The converse implies that the less-bound stars, despite being hard relative to the cluster core energy, are not hard relative to the more tightly bound companion to the IMBH. These less bound stars exchange energy with the innermost binary in a manner which prevents their orbits from tightening significantly. 
Because strong perturbations drive the binary hardening process, the continuous hardening of the most-bound star's orbit can be approximated  relative to the properties of the cluster background as 
\beq
{d a \over dt } \propto {  G \rho_{\rm c}  \over \sigma_{\rm c}}  a_0^2,
\eeq
where  $\rho_{\rm c}$ is the mass density of the cluster core environment surrounding the binary \citep{1996NewA....1...35Q,2003gmbp.book.....H}.  This hardening rate implies that the timescale for change of the semi-major axis, $a_0/\dot a_0$ scales as $a_0^{-1}$ because $\dot a_0 \propto a_0^2$.  These scalings for the time-evolution of the most-bound object's semi-major axis suggest the qualitative behavior seen in Figure \ref{fig:radii_ks}, where $a_{0}$  first decreases rapidly while $a_0$ is large, then more slowly as $a_0$ decreases.

\subsection{Companion Orbital Properties}

\begin{figure}[tbp]
\begin{center}
\includegraphics[width=0.49\textwidth]{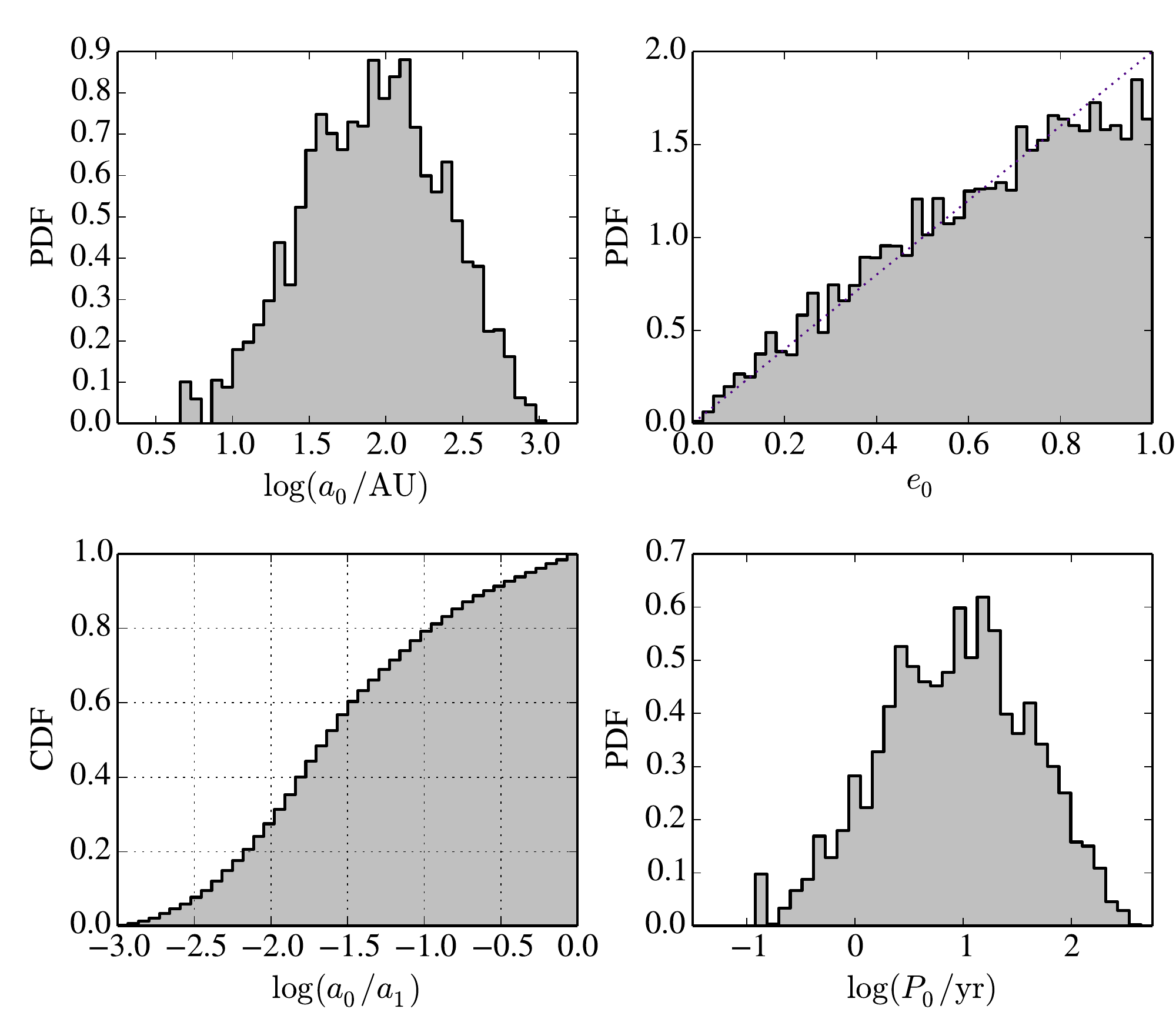}
\caption{Orbital properties of closest companions to the IMBH sampled at equal time intervals.  
We plot semi major axis, $a_0$, eccentricity, $e_0$, and period $P_0$, of the most-bound companion. We also show the ratio, $a_0/a_1$, which describes the compactness of the innermost companion's orbit compared to the 2nd most-bound star. The  $a_0/a_1$ distribution shows that the IMBH has a close companion that is significantly more tightly bound than any other objects a large fraction of the time. Approximately 90\% of the time,   $a_0/a_1 < 1/3$. 25\% of the time, the companion is $100 \times$ more tightly bound than any other star. Typical orbital semi-major axes range from a few to a thousand AU,  and the period distribution is also broad, peaking around 10~yr.  The eccentricity distribution is nearly thermal, with some a slight decrement arising at $e_0 \sim 1$  due to the possibility of tidal disruption by the IMBH. 
}
\label{fig:comp_orb_prop}
\end{center}
\end{figure}

In this section, we examine some of the orbital properties that define the population of most-bound companions to the IMBH. 
Figure \ref{fig:comp_orb_prop} highlights some of the key orbital demographics of these companions.  
The lower left panel of Figure \ref{fig:comp_orb_prop}  displays the cumulative distribution function of the log of ratio of the semi-major axis of the most-bound star to the second most bound $\log ( a_0/a_1)$.  This diagram reveals that, in our simulations, the IMBH has a companion that is hierarchically isolated from the rest of the cluster stars a large fraction of the time. 90\% of the time, the companion has $a_0 < {1\over3} a_1$. Approximately 25\% of the time, $a_0 < 10^{-2} a_1$.

The other panels of Figure  \ref{fig:comp_orb_prop} examine the distributions of orbital characteristics in physical units. In the upper left panel, we show the orbital semi-major axis distribution of most-bound companions.  The largest separations are around $10^3$ AU. This distance becomes similar to the average distance of the less-bound stars within the IMBH's sphere of influence (see, for example Figure \ref{fig:radii_ks}), and, as a result, determines the maximum separation at which an object is able to persist as most-bound without exchange. The peak of the distribution, lies in the range of $10^1 - 10^2$ AU, and the minimum separation observed in our simulations is a few AU. The upper right panel shows the eccentricity distribution of companion objects. It is nearly thermal for most of the range of $e$, with some decrement observed near $e\sim1$, because objects that enter into orbits that are too eccentric interact directly with the IMBH at orbital pericenter. The orbital separation distribution, along with the object masses, map to an orbital period distribution, shown in the lower right panel of Figure \ref{fig:comp_orb_prop}.

\begin{figure}[tbp]
\begin{center}
\includegraphics[width=0.49\textwidth]{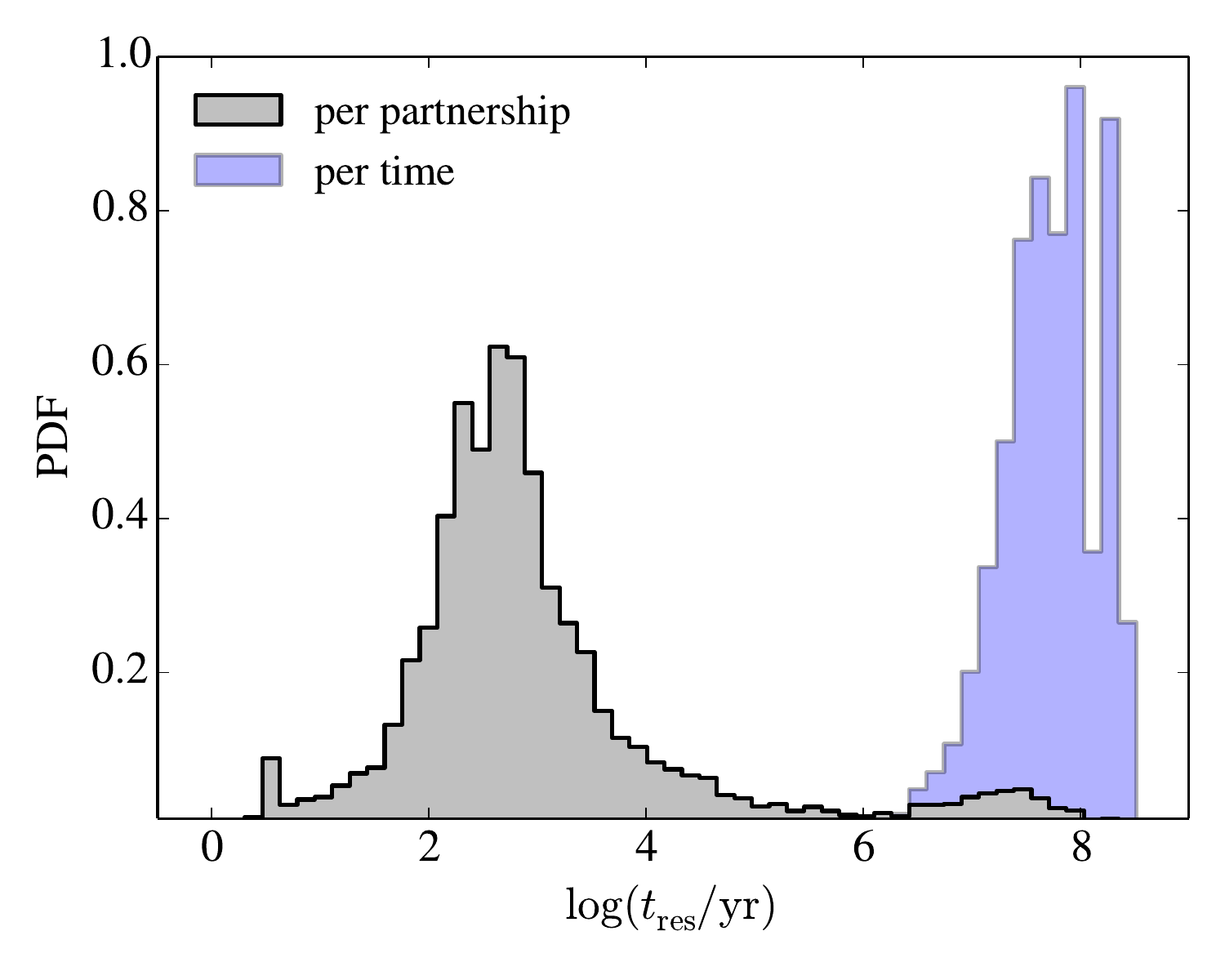}
\caption{Residence time of closest-partner status for companions to the IMBH. The grey distribution shows the time span of each individual closest-partnership formed with the IMBH  (samples {\it per unit partnership}). The blue distribution samples residence time of the IMBH's parter at equal time intervals  (samples {\it per unit time}). Thus the grey distribution shows how long partnerships last, while the blue distribution shows the residence time likely to be seen by an observer seeing the cluster at a random time.  An examination of the grey distribution shows that many partnerships with the IMBH suffer a breakup after $\sim 10^3$ yr, a timescale associated with one to several orbital periods. A tail of longer-residence time objects exists, though, followed by a secondary maximum at timescales around $10^7$ yr, representative of the objects that hierarchically isolate from the remaining cluster stars. The blue distribution shows that these long-lasting partnerships form a dominant contribution to the typically observed IMBH companions. This qualitative conclusion can be inferred visually from Figure \ref{fig:radii_ks}, which shows shows that the IMBH typically has a long-lasting companion (with residence time $\gtrsim10^6$ yr). }
\label{fig:res}
\end{center}
\end{figure}

As can be inferred from an inspection of Figure \ref{fig:radii_ks}, companions to the IMBH do not persist indefinitely. Instead these objects undergo cycles of replacement following their termination -- either through exchange or direct interaction with the IMBH. 
In Figure \ref{fig:res}, we show the residence time of a given object as the companion to the IMBH. In the grey histogram, we measure the persistence time of each individual partnership with the IMBH.  With this diagnostic, we can examine how long a typical close partnership between the IMBH and a companion lasts. Figure \ref{fig:res} reveals a bimodal distribution of persistence times of partnerships between the IMBH and other objects. One peak of this distribution lies around $10^2-10^4$ yr, the second peaks around $10^7$ yr.  A  tail toward longer residence times connects the two peaks.  The origin of the first peak in this distribution can be traced back directly to the orbital period of the typical most-bound companion, as show in the right-hand panel of Figure \ref{fig:comp_orb_prop}. The longest orbital periods of the most-bound companion occur when $a_0\sim a_1$, and are $10^2$ yr for the parameters used here. The short residence time branch can thus be associated with objects that persist as partner to the IMBH for only one orbit. 
Near the peak of the distribution, the grey histogram in Figure \ref{fig:res} records objects that last for a small integer number of orbits. Associating the peak at $\sim 10^3$ yr in Figure \ref{fig:res} with $N$ orbits of of an object near the maximum period for most-bound companions, we find that $\sim 10$ orbits is a typical residence time before an exchange of hierarchy.  A  tail to longer residence times represents objects whose orbits harden and thus persist for many more orbital periods.  

The origin of the secondary peak at $\sim 10^6 - 10^8$ yr arises from objects whose orbits harden sufficiently to become significantly hierarchically isolated from the remaining cluster.
 This timescale thus becomes comparable to the cluster relaxation timescale of $\sim 0.6$~Gyr.   In the blue histogram of Figure \ref{fig:res}, we show the distribution of residence times for most-bound companion objects sampled at even time intervals  (therefore it is possible to count the same long-lived companion at multiple sampling times in this case). If we observe the cluster at any given time, we are very likely to see a companion with residence time of $10^6 - 10^8$~yr. This impression is confirmed visually in Figure \ref{fig:radii_ks}, where the populations of objects that we see in the timeseries last, on average, for timescales of fractions of a Gyr or more.

\subsection{Companion Stellar Properties}

The color coding of stellar evolutionary state in Figure \ref{fig:radii_ks} shows that the demographics of the most-bound objects are remarkably diverse.  This outcome is similar to the results from \citet{2014ApJ...794....7M} for systems undergoing Roche Lobe overflow. Companion objects include BHs (1-2 Gyr), as well as MS and giant-branch stars, and WDs of various masses. In some cases, a single object will persist as the closest companion for long enough that it evolves and its stellar evolutionary state changes. One example of this is seen between $\sim 2.2-2.4$ Gyr, where a single most-bound star evolves from the MS to the giant branch, and eventually tidally interacts with the IMBH. In this section, we examine the statistical distributions of stellar evolutionary type, mass, and radius of the most-bound companion objects.

\begin{figure}[tbp]
\begin{center}
\includegraphics[width=0.49\textwidth]{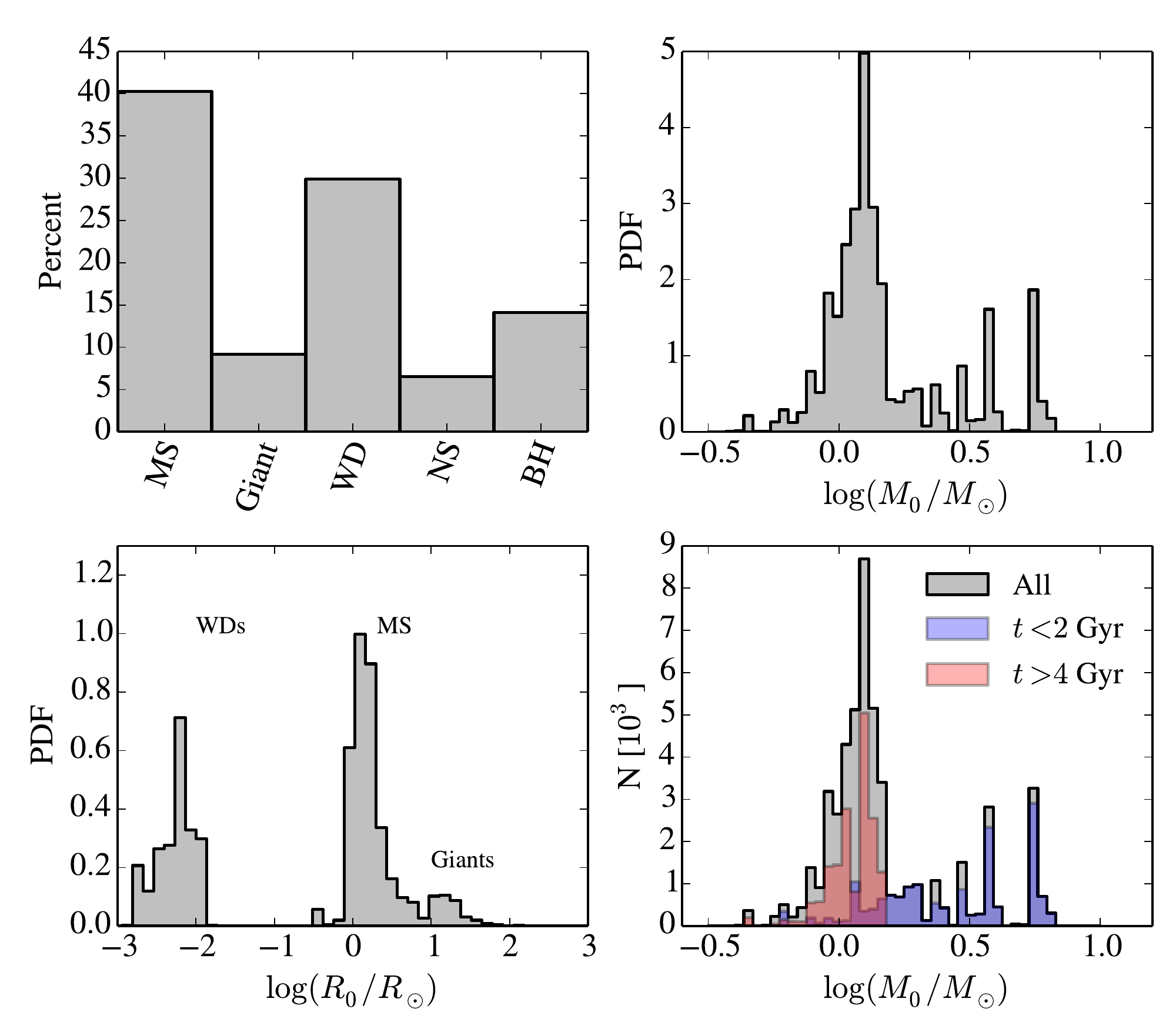}
\caption{Stellar properties of closest companions to the IMBH sampled at equal time intervals. 
The upper-left panel shows companion stellar evolutionary type (for a key see the legend of Figure \ref{fig:radii_ks}), the upper-right panel shows companion masses $M_0$, and the lower-left panel their radii, $R_0$.  Companions include comparable fractions of stars (both MS and giants) and stellar remnants (WDs, NSs, and BHs). The typical mass of BH companions is substantially higher than the initial cluster mean mass of $0.58M_\odot$. However, a broad distribution suggests that even relatively low mass companions $\sim 0.3 M_\odot$ occasionally find themselves tightly bound to the IMBH. The lower-right panel shows how the mass distribution evolves in time. At early times in the simulation, shown in blue, the mass distribution is weighted toward high masses, and includes both massive stars and massive stellar remnants. Late times, when the cluster turnoff mass is lower, contribute substantially to the lower-mass portion of the distribution.  }
\label{fig:comp_stellar_prop}
\end{center}
\end{figure}

The diversity of companion types seen in Figure \ref{fig:radii_ks} is reflected quantitatively in Figure \ref{fig:comp_stellar_prop}. 
The upper left panel of this figure shows the companions' stellar evolutionary type. The most prevalent companions to the IMBH are MS stars, but compact objects like WDs, NSs, and BHs all represent significant contributions to the total demographics. The companion's radius distribution, which we show in the lower left panel of Figure \ref{fig:comp_stellar_prop}, directly reflects the stellar evolutionary status of the companions. We show the range of radii encompassing WDs to giant stars in this panel. The primary peaks of the radius distribution lie at $10^{-2} R_\odot$, and and $\gtrsim 1 R_\odot$, reflecting the WD and massive MS populations, a smaller third peak is seen between $10^{1} - 10^{2} R_\odot$ that includes stars at various phases along the giant branch.

The objects' mass distribution, shown in the right hand panels of Figure \ref{fig:comp_stellar_prop}, exhibits a broad main peak centered at slightly more than a solar mass that is made up of MS stars, giants, massive WDs, and NSs.  This peak lies substantially above the mean mass of cluster stars, which is initially $\sim 0.58 M_\odot$. We also observe a tail toward much higher companion masses (up to $\sim 10 M_\odot$) made up mostly of MS stars, giants, and stellar-mass BHs.  That the mass of most of the IMBH's companions is significantly larger than that of a typical cluster star is not surprising. The trend toward energy equipartition in two-body interactions causes stars more massive than the mean mass to sink toward orbits deep in the cluster core, where they are very likely to come into direct interaction (and potentially partnership) with the IMBH \citep[e.g.][]{2012ApJ...752...43G,2004ApJ...604..632G,2014MNRAS.444...29L}. Further, more massive secondary objects are likely to replace an existing companion if they do undergo a three-body interaction \citep{1982AJ.....87..175F,1993ApJ...415..631S,2014ApJ...784...71S,2014ApJ...794....7M}. In the lower right panel of Figure \ref{fig:comp_stellar_prop} we show the un-normalized distribution of companion masses in order to demonstrate that there is strong evolution in the companion mass distribution with stellar population age. Young ages,  $<2$~Gyr, where the turnoff mass is still high, contribute the bulk of the companions $\gg 1 M_\odot$. 
This temporal dependence in the companion mass can be contrasted to the approximately time-independent companion orbital parameter distributions described in the previous sub-section.

\begin{table}[tbp]
\begin{center}
\begin{tabular}{cccccc}
\tableline\tableline
Name & MS & Giant  &  WD & NS & BH\\
 & [\%] & [\%] & [\%]& [\%]& [\%] \\
\tableline
A & 40.3 &  9.2 &   29.9 &   6.5 &  14.1  \\
B & 11.5 & 2.3 &   12.5 &    6.4 &   67.3  \\
C & 27.6 & 6.0 &   49.6 &   1.6 &  15.2  \\
D & 21.9 &  3.2 &  43.6 &   4.0 &  27.4  \\
\tableline
\end{tabular}
\caption{Demographics of stellar type of the most-bound companions to the IMBH. Tabulated by percentage as sampled at equal time intervals. }\label{table:demo}
\end{center}
\end{table}

\subsection{Termination of Close Partnerships}

Close partnerships to the IMBH can be terminated through one of several channels. In this section, we describe the possible channels and explore their relative likelihoods.

\subsubsection{Channels}

\begin{figure}[tbp]
\begin{center}
\includegraphics[width=0.49\textwidth]{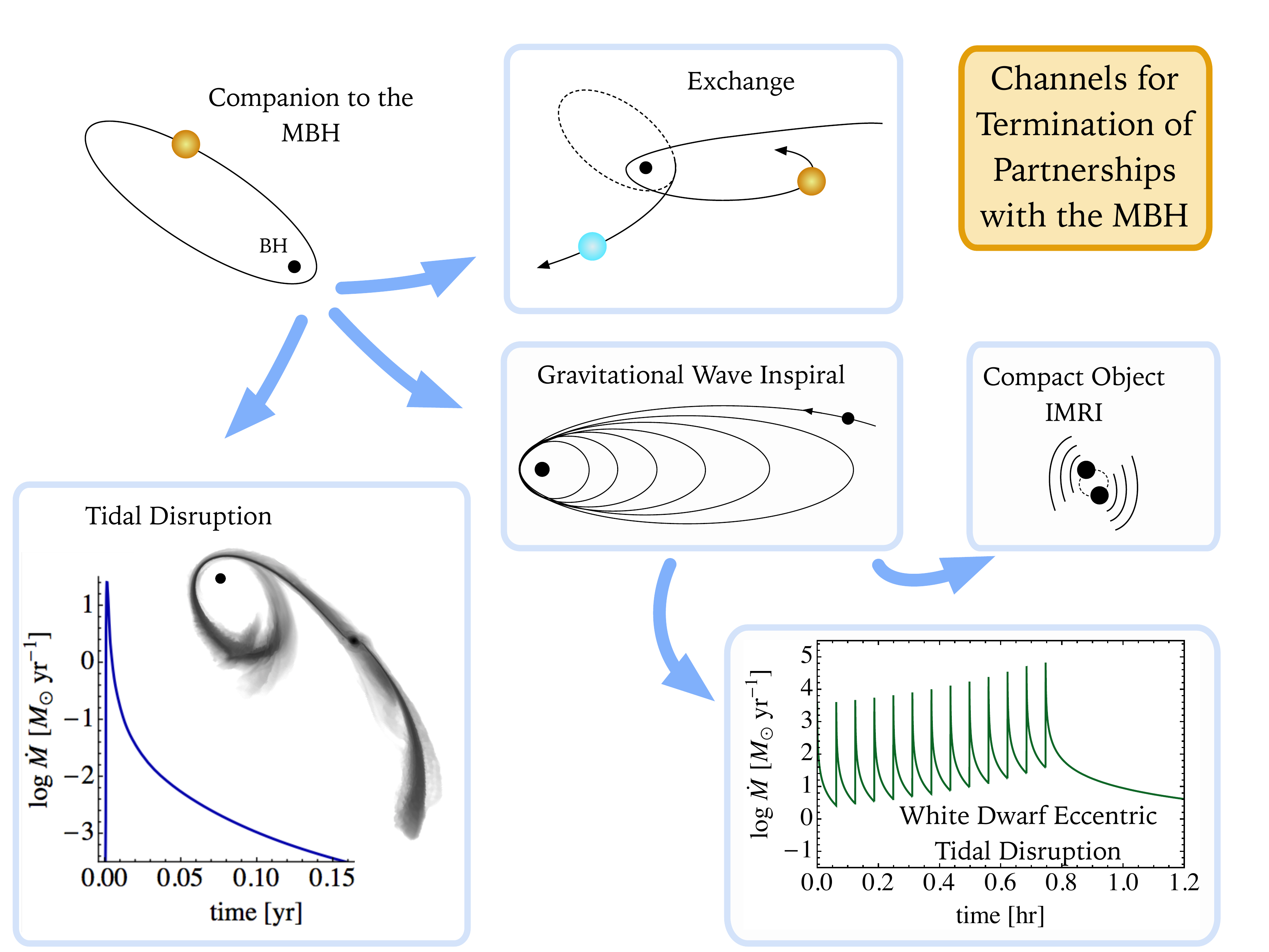}
\caption{ Interaction channels that can bring about the end of a close partnership between the IMBH and another object. Possible outcomes of a close partnership include three and multi-body exchanges, gravitational wave inspirals and tidal disruptions. The bulk of partnerships are terminated through the exchange channel, but a fraction of long-lasting binary companions may end up tidally disrupted or undergoing an inspiral. If the inspiralling object is a BH or NS, the initial gravitational wave capture leads to an IMRI \citep[][]{2007CQGra..24R.113A,2008ApJ...681.1431M,2009CQGra..26i4036M,2013A&A...557A.135K}, while if the object is a WD, tidal heating of the object likely leads to disruption while from an eccentric orbit and repeated flaring episodes \citep{2014ApJ...794....9M}.  }
\label{fig:diagram}
\end{center}
\end{figure}

Three possible channels of termination of a close partnership with the IMBH exist. They are shown in the diagram of Figure \ref{fig:diagram} and explained below. 

\begin{itemize}
\item {\bf Exchange:}
Exchanges can occur when a third body's pericenter distance from the IMBH becomes similar to the semi-major axis of the binary companion $a_0$.  When this occurs, the system undergoes a strong three-body interaction, during which the gravitational attraction between both objects and the IMBH is similar. Either star may emerge from the three-body interaction as the new most-bound object. Statistically, there is also a chance that the IMBH could be the ejected object, leaving behind a binary of two stars. However, since the mass ratio is large, this exchange outcome is extremely unlikely \citep{1996ApJ...467..359H}.   Throughout the duration of our simulations, we record all perturbers to the central binary with pericenter distance less than 3 times $a_0$. This allows us to track the statistics of these exchange interactions.  We find that the bulk of exchanges are due to strong three-body interactions, where the previous perturber emerges from the encounter as the new companion.  A small fraction, of order a few percent, are the result of more complex multi-body strong interactions where an object other than the strongest perturber emerges as the companion immediately following the encounter.

\item {\bf Tidal Disruption:}
An examination of Figure \ref{fig:radii_ks} reveals that despite having the highest binding energy to the IMBH, the orbit of the most-bound star is still subject to perturbations.  Not only does its orbit tighten through interactions with other cluster-core stars, but its orbital angular momentum wanders in magnitude and direction, as can be seen through the orbital pericenter distance, $r_{\rm p,0}$. As a result, the most-bound star will be occasionally perturbed into an orbit which leads it to pass within a tidal radius, $\rt$, of the IMBH.  In our simulations, we record these tidal disruption events and delete the particle from the simulation, as described in Section \ref{sec:method}. In reality, some stars, especially those with condensed cores, can lose part of their envelope gas without being completely disrupted \citep{2012ApJ...757..134M,2013ApJ...777..133M}.  The remnants of these partial tidal stripping events could then go on to further interaction with the IMBH \citep{2005MNRAS.363L..56H,2014ApJ...788...99B}.

\item {\bf Gravitational Wave Inspiral:}
Just as the orbital wandering of the most-bound object permits the occasional tidal disruption event, if the companion to the IMBH is a compact object it may also undergo a intermediate mass ratio gravitational wave inspiral  (IMRI) \citep{2007CQGra..24R.113A,2008ApJ...681.1431M,2008MNRAS.391..718S,2009CQGra..26i4036M,2013A&A...557A.135K}. In our simulations, these are flagged as occurring when the gravitational wave inspiral time becomes less than one $N$-body time unit, and the particle is removed from the simulation (see Section \ref{sec:method}). \citet{2008ApJ...686..829H} and \citet{2013A&A...557A.135K} explore the retention of IMBHs suffering a gravitational wave recoil following an IMRI event. IMBHs that merge with stellar mass BHs with mass ratio $q \gtrsim 0.1$ have a significant probability of being ejected from the cluster. Smaller kicks would temporarily offset the IMBH from the cluster core. 
These inspirals typically occur when the companion is in a very eccentric orbit, and are driven primarily by the close pericenter distance, rather than, for example a close semi-major axis, much like the tidal disruptions \citep[and in agreement with the findings of][]{2004ApJ...616..221G}.  The $N$-body simulations of \citet{2004ApJ...613.1143B} and \citet{2014MNRAS.444...29L} lack this merger channel and \citet{2014MNRAS.444...29L} demonstrate that this leads to unrealistically long-lived central partnerships.  If the companion is a BH or NS it will be swallowed as it merges with the IMBH. However, in some cases where the companion is a WD, it will tidally disrupt before merging with the IMBH \citep[e.g.][]{2006MNRAS.365..929C,2007A&A...476..121I,2008MNRAS.391..718S,2010MNRAS.409L..25Z,2011ApJ...743..134K,2014ApJ...794....9M}. This can proceed in one of two ways. In most cases, tidal heating likely leads to the disruption of the WD while it is still in an eccentric orbit. However, in some cases the orbit may fully circularize before the onset of Roche Lobe overflow. 

\end{itemize}

\subsubsection{Termination Demographics}

\begin{figure*}[tbp]
\begin{center}
\includegraphics[width=0.99\textwidth]{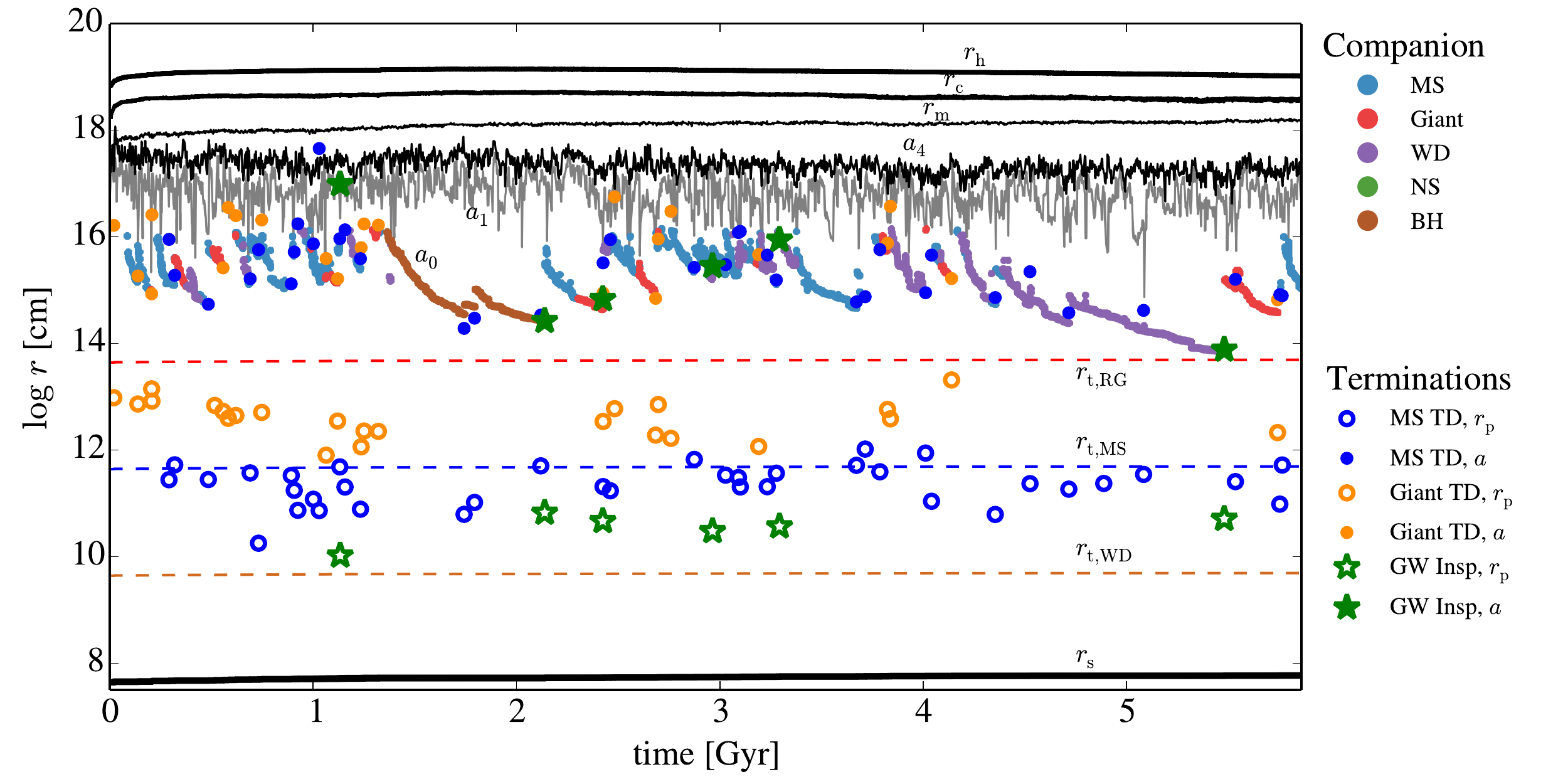}
\caption{Characteristic radii measured with respect to the IMBH in the same example timeseries as Figure \ref{fig:radii_ks}, from simulation group A. Here we expand the domain to include the IMBH's Schwarzschild radius, $r_{\rm s}$, and estimated tidal radii of typical WDs (brown), MS stars (blue), and giants (red). Overset on this timeseries we plot the semi-major axis (solid symbols) and pericenter distance (open symbols) of tidal disruption events (circular points) and gravitational wave inspirals (stars). These direct interactions with the IMBH often lead to the termination of long-standing partnerships with the IMBH. On visual inspection, several categories of events emerge. Gravitational inspirals are particularly likely to come at the end of a long partnership when the orbital SMA has decreased. Tidal disruptions, by contrast, often come in times of multi-body interaction that scatter stars toward the IMBH. These are particularly likely to occur when $a_0 \sim a_1$, either because $a_0$ is large or $a_1$ is smaller than average. In several cases represented here a  most-bound star evolves off the MS and  up the giant branch until it eventually reaches the point of tidal disruption due to its growing radius. 
}
\label{fig:radii_dis}
\end{center}
\end{figure*}

\begin{table}[tbp]
\begin{center}
\begin{tabular}{ccccccc}
\tableline\tableline
Name & $\dot N_{\rm ex}$ & $\dot N_{\rm insp}$ & $\dot N_{\rm tde}$ & $f_{\rm ex,3b}$ & $f_{\rm insp,0}$ & $f_{\rm tde,0}$ \\
 & [Gyr$^{-1}$] & [Gyr$^{-1}$] & [Gyr$^{-1}$] & [\%] & [\%] &  [\%] \\
\tableline
A & 530.86 & 1.77 & 10.11 & 98.62 & 80.65 & 43.50 \\
B & 253.01 & 1.76 & 4.54 & 98.32 & 83.87 & 33.75 \\
C & 318.70 & 1.76 & 5.22 & 99.16 & 83.33 & 47.89 \\
D & 884.06 & 3.65 & 11.68 & 98.78 & 78.95 & 30.14 \\
\tableline
\end{tabular}
\caption{Close encounter and termination Demographics of most-bound companions to the IMBH.  }\label{table:term}
\end{center}
\end{table}

The termination of partnerships with the IMBH, as well as other close encounters, are displayed in Figure \ref{fig:radii_dis}. In this figure, we show the same timeseries plotted in Figure \ref{fig:radii_ks}. Here, we extend the range of the plot include radii very close the the IMBH, including the IMBH's Schwarzschild radius, $r_{\rm s}$. We also include tidal radii of typical WDs, MS stars and giant stars. Overset on this timeseries we highlight the orbital semi-major axis (solid symbol) and pericenter distance (open symbol) of encounters leading to either tidal disruption or gravitational wave inspiral.  Tidal disruptions are shown with circular points colored blue for MS star disruptions and red for giant star disruptions.  Gravitational wave inspirals are marked with green stars. 

Figure \ref{fig:radii_dis} reveals the broad demographics of changes of partnership with the IMBH. Many changes of partnership appear to arise from strong interactions between the companion to the IMBH and other cluster stars -- often including the second most-bound star, whose semi-major axis, $a_1$, is occasionally similar to $a_0$. When $a_0\sim a_1$ changes of partnership are frequent, and as a result the constantly-changing companion's orbit does not have the opportunity tighten significantly. An example of this scenario is seen in the $2.5 - 4.5$ Gyr span of Figure \ref{fig:radii_dis}.  Close encounters with the IMBH often accompany strong encounters -- many tidal disruption events either involve the most-bound companion or stars scattered into a similar orbit. The chaotic three-body dynamics of the IMBH, companion, and perturber system then cause a fraction of stars to be scattered directly toward the IMBH.   An exception to this requirement for scatterings comes in the form of giant star disruptions, which can occasionally occur when the star evolves to become tidally vulnerable to the IMBH rather than being strongly scattered \citep[`spoon-feeding' the IMBH:][]{2013ApJ...777..133M}.

In Table \ref{table:term}, we examine the termination of close partnerships with the IMBH as well as the rates of close encounters with the IMBH.  The most common, by far, means for changes of partnership with the IMBH is the exchange interaction, contributing approximately 98\% of the changes in partnership. These exchanges dominate the large peak of IMBH companions that have residence times in the range of $1-10^6$ years, as shown in Figure \ref{fig:res}.  The mean exchange rate is  $2.5-5\times10^{-7}$ yr$^{-1}$.  Of these exchanges, $\sim$98\% are simple three-body exchanges in which the strongest perturber becomes the the new most-bound companion. The remaining 2\% are more complex multi-body interactions. These may involve more than three objects, or they may involve the tidal disruption or inspiral of one or more objects. Gravitational wave inspirals and tidal disruption events contribute to the remaining termination of partnerships with the IMBH. The tidal disruption rate is of order $10^{-8}$ yr$^{-1}$ and the inspiral rate is of order $10^{-9}$ yr$^{-1}$ in our simulations. In Section \ref{sec:disc}, we compare these event rates to previous predictions and mention their potential significance for electromagnetic and gravitational transients.

As can be inferred from Figure \ref{fig:radii_dis}, the longer a companion object persists, the more likely it is that the pair will be split by direct interaction with the IMBH rather than exchange. Longer lived companion pairs segregate to smaller orbital semi-major axes, reducing the cross-section for encounters or exchange in proportion to $a_0$.  Meanwhile, interaction with the IMBH becomes progressively more likely at smaller $a_0$ because the orbital eccentricity needed to produce a close pericenter passage becomes smaller.  Thus, many long-lived companions appear to only be replaced when they are scattered to their disruption, or when they undergo a gravitational wave inspiral. These terminations contribute to the large fraction of tidal disruption events and gravitational wave inspirals arising from the most-bound companion in Table \ref{table:term}.

%
% Cluster parameters
%
\section{Dependence on Cluster Properties}\label{sec:diff}

\begin{figure*}[tbp]
\begin{center}
\includegraphics[width=0.99\textwidth]{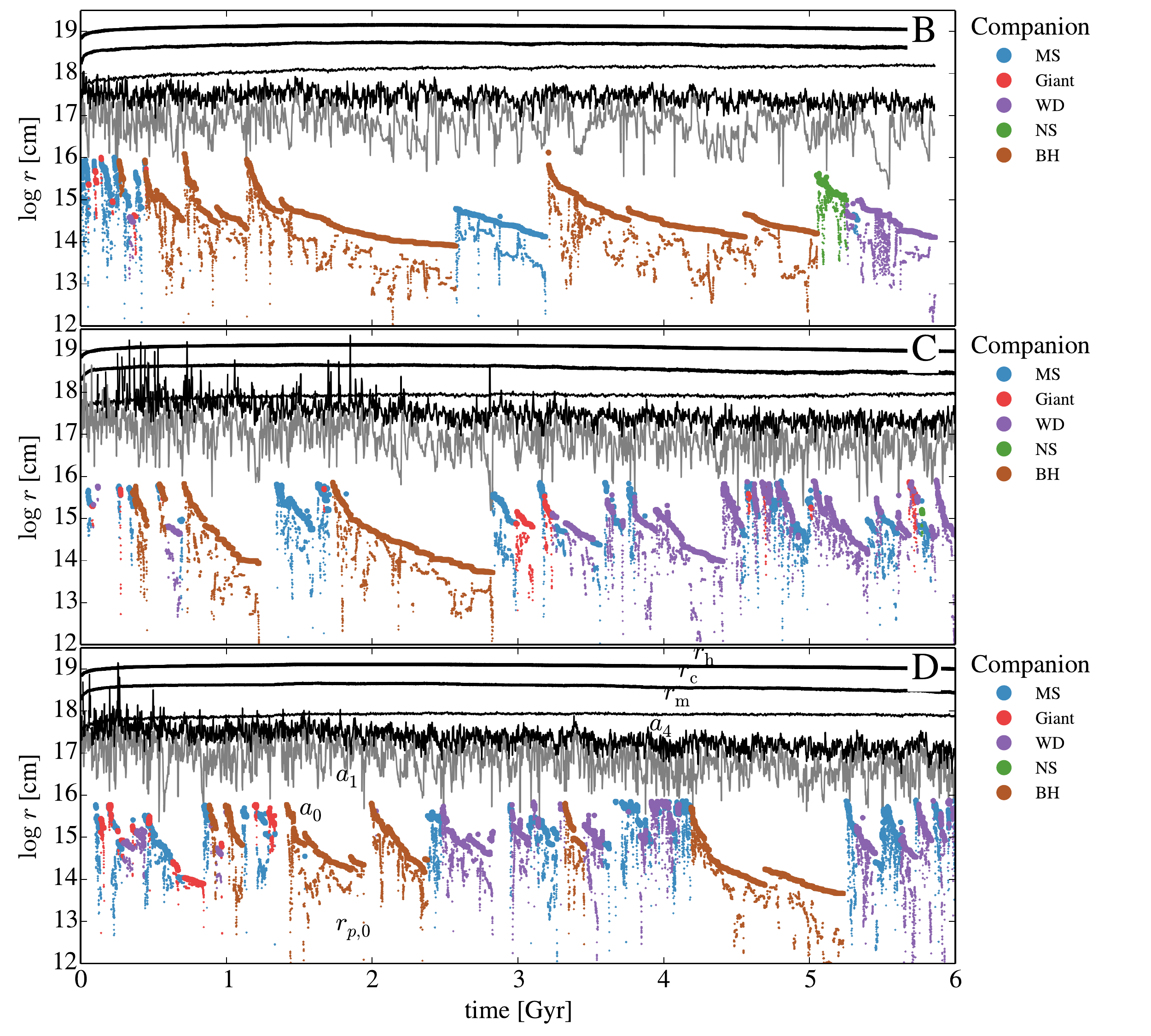}
\caption{Comparative timeseries of first 6 Gyr of cluster evolution of examples from simulation groups B-D. This figure is analogous to Figure \ref{fig:radii_ks}, which plots the evolution of one of the simulation group A models. As in Figure \ref{fig:radii_ks}, we color hard-binary companions to the IMBH based on their evolutionary type.  Table \ref{table:sims} describes the differences between simulation groups B-D as compared to A. }
\label{fig:simcomp_timeseries}
\end{center}
\end{figure*}

Having outlined the basic process of IMBH companion formation, evolution, and destruction through reference to simulation group A in the previous section, here
 we consider the dependence of our results on the chosen simulation parameters. We run several sets of simulations A-D in Table \ref{table:sims}, which allow us to systematically explore the differences in the properties of close companions to the IMBH with varying cluster properties. 
 These variations prove useful in understanding the physics underlying companion object dynamics. 
 Simulation group B has a lower kick velocity dispersion that our fiducial case, A, which we have up until now analyzed. This lower kick velocity allows a larger fraction of remnant NSs and BHs to be retained in or near the cluster core. Simulation group C has a lower IMBH mass than the fiducial case by a factor of two -- it's initial mass is $75M_\odot$ rather than $150 M_\odot$. Finally, simulation group D has a larger total number of stars (and mass) by a factor of 2 ($2\times10^5$ rather than $10^5$). By exploring these systematic differences, we can also consider the extrapolation of our results to higher mass clusters or IMBHs.

\subsection{Companion Objects in Cluster Evolution}

In Figure \ref{fig:simcomp_timeseries}, we plot comparative timeseries, analogous to that shown in Figure \ref{fig:radii_ks}, for examples from simulation groups B-D (which may be compared directly to Figure \ref{fig:radii_ks} for simulation group A). Even looking at the examples in this figure, dynamical and demographic differences and similarities distinguish themselves. The pattern of orbital tightening and replacement of a most-bound companion is repeated throughout the panels, despite their varying cluster and IMBH parameters. The global evolution of the clusters is largely similar as well: cluster core and half mass radii are relatively constant in time, just as with simulation group A. 

Differences in the stellar dynamics and demographics of the companion objects  also become apparent on closer inspection of Figure \ref{fig:simcomp_timeseries}. Simulation group B, with a larger population of retained BHs, shows a larger quantity of BHs joining the IMBH as most-bound companions. 
The cycles of orbital hardening experienced by these BHs are longer-lived than those of most of their stellar counterparts. By contrast, simulation group D, with a larger total number of stars than the fiducial case shows more frequent exchanges of partnership, but a diversity of companion-object stellar evolutionary types.

\begin{figure}[htbp]
\begin{center}
\includegraphics[width=0.43\textwidth]{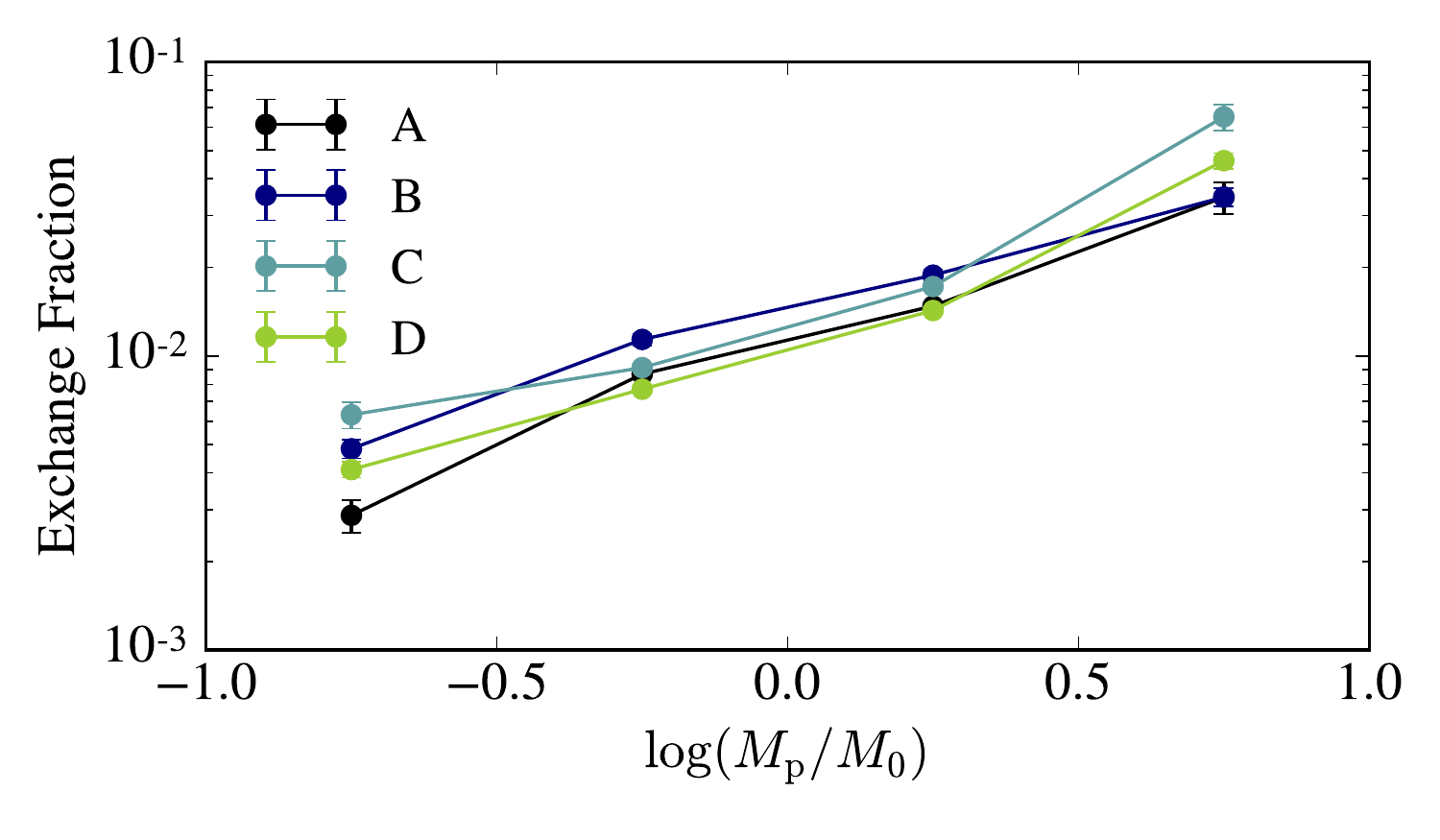}
\caption{Fraction of close encounters between perturbers of mass $M_{\rm p}$ and the IMBH and companion pair that result in an exchange. The exchange fraction depends sensitively on the perturber to companion mass ratio, allowing massive objects to exchange into the inner binary easily, and then to resist being exchanged out by subsequent encounters with lighter objects. By contrast, we see little dependence of the exchange fraction on other cluster properties, like the IMBH mass.  Errorbars are poisson errors on individual bins in the exchange cross section.  }
\label{fig:ex}
\end{center}
\end{figure}

These BHs preferentially enter into partnerships with the IMBH because of their significantly higher-than-average mass, which causes them to sink in the cluster potential due to dynamical friction. 
When these stellar-remnant BHs interact with the IMBH and its companion, their high mass makes them very likely to replace an existing companion. \citet{1996ApJ...467..359H} has shown that the exchange cross section depends quite strongly on the mass of the tertiary, perturbing body, even in the case where the mass ratio between the binary components is large. 
In Figure \ref{fig:ex}, we plot the fraction of encounters that result in an exchange given a particular perturber to companion mass ratio, $M_{\rm p} / M_0$. This fraction is computed by counting the fraction of encounters in which the perturber comes within $a_0$ of the binary center of mass at pericenter that result in an exchange. While the exchange fraction depends strongly on perturber mass ratio, it does not depend nearly as strongly on other cluster properties.  This plot implies that massive objects can exchange into the inner binary with relative ease and that, as companions, they can undergo many encounters, each hardening their orbit before one leads to an exchange.

\subsection{Companion Demographics}

\begin{figure}[tbp]
\begin{center}
\includegraphics[width=0.49\textwidth]{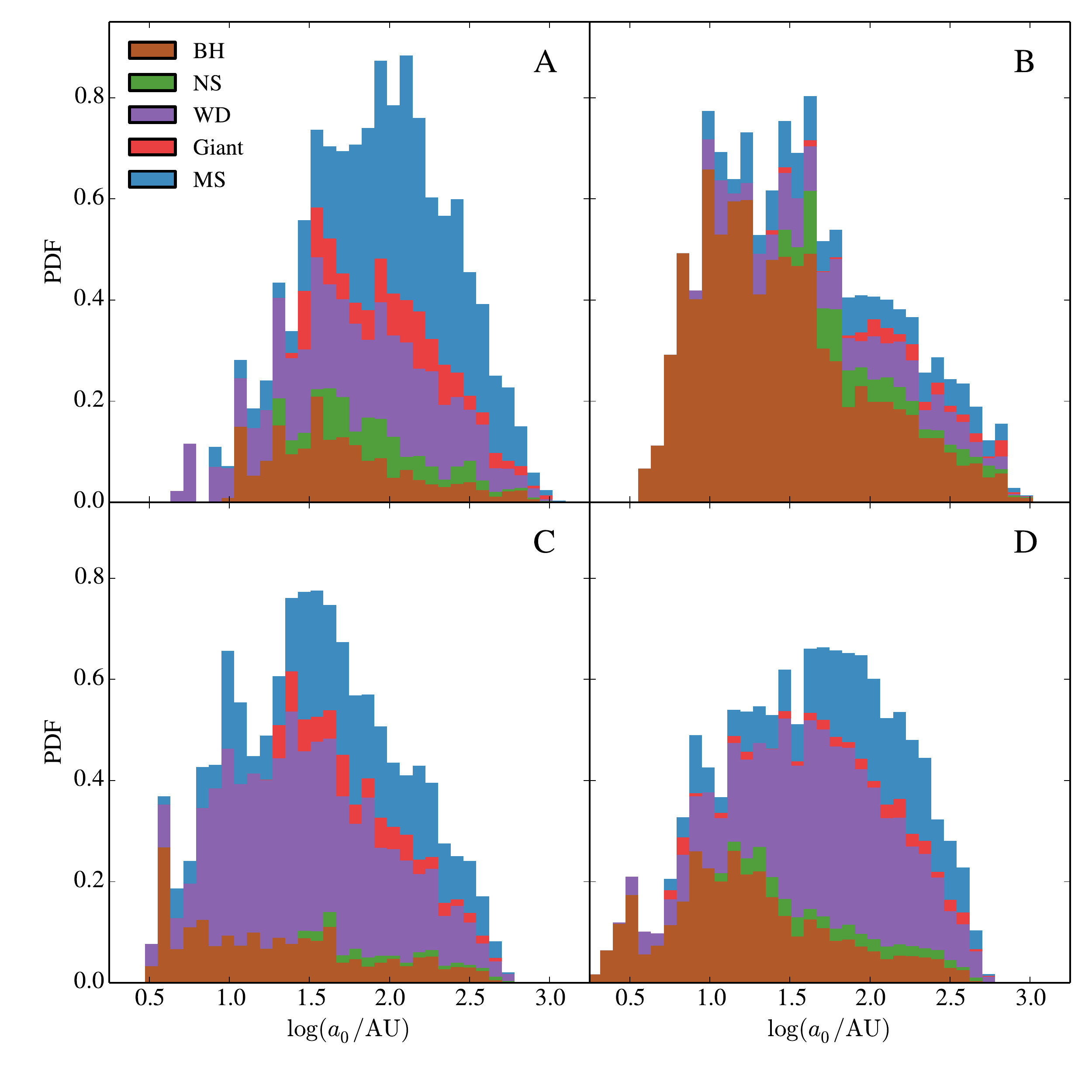}
\caption{Distributions of companion object semi-major axis, $a_0$, subdivided to show companion demographics. Panels show simulation groups A-D outlined in Table \ref{table:sims}.  These panels illustrate how differing companion demographics between simulation groups propagate to different distributions of companion orbital properties. The distributions of compact object (BH and WD especially) semi-major axes peak at smaller values than the overall distribution, so varying quantities of compact object companions propagate to different overall distributions of companion orbits.   }
\label{fig:smademo}
\end{center}
\end{figure}

Figure \ref{fig:smademo} further traces how companion demographics propagate to orbital distributions. The histogram of orbital semi-major axis is shown for each of the simulation groups A-D, and is broken down by companion type. Visually from Figure \ref{fig:simcomp_timeseries}, and here quantitatively, we can infer that types of companions have different sub-distributions. When combined in varying percentages, these demographic differences lead to variations in the overall semi-major axis distribution of IMBH companions.  The most obvious example are BHs. Based on the histograms of Figure \ref{fig:smademo}, BHs exist in orbits which are preferentially among the most tightly bound. The distribution is broad and over the course of the simulation there were BHs at all separations. We can understand this distinction because BHs are preferentially difficult to replace in exchange interactions due to their high mass, and are not prone to tidal disruption when the the companion has orbit with $e\sim 1$. 

By contrast, giant stars tend to exist in the least-bound orbits in Figure \ref{fig:smademo}. These objects are especially tidally vulnerable, and so are subject to tidal disruption by the IMBH at larger orbital separations for a given eccentricity than are more compact MS stars or WDs. 
The cases of MS stars and WDs lie intermediate to that of giants and  BHs. MS stars' innermost semi-major axes are also limited by tidal disruption as their orbits cycle through high-eccentricity phases (note the rapidly varying eccentricity of the innermost orbit in Figure \ref{fig:radii_ks}). WDs are less tidally vulnerable, but are not as massive as the BHs, so their distribution peaks at somewhat larger separations. The case of the NSs is interesting and somewhat more subtle. Although these remnants are $\sim 3 \times$ the average stellar mass in the cluster, they are subject to kicks on formation and they do not re-segregate to the cluster core as efficiently as the more-massive BHs.

\begin{figure}[tbp]
\begin{center}
\includegraphics[width=0.49\textwidth]{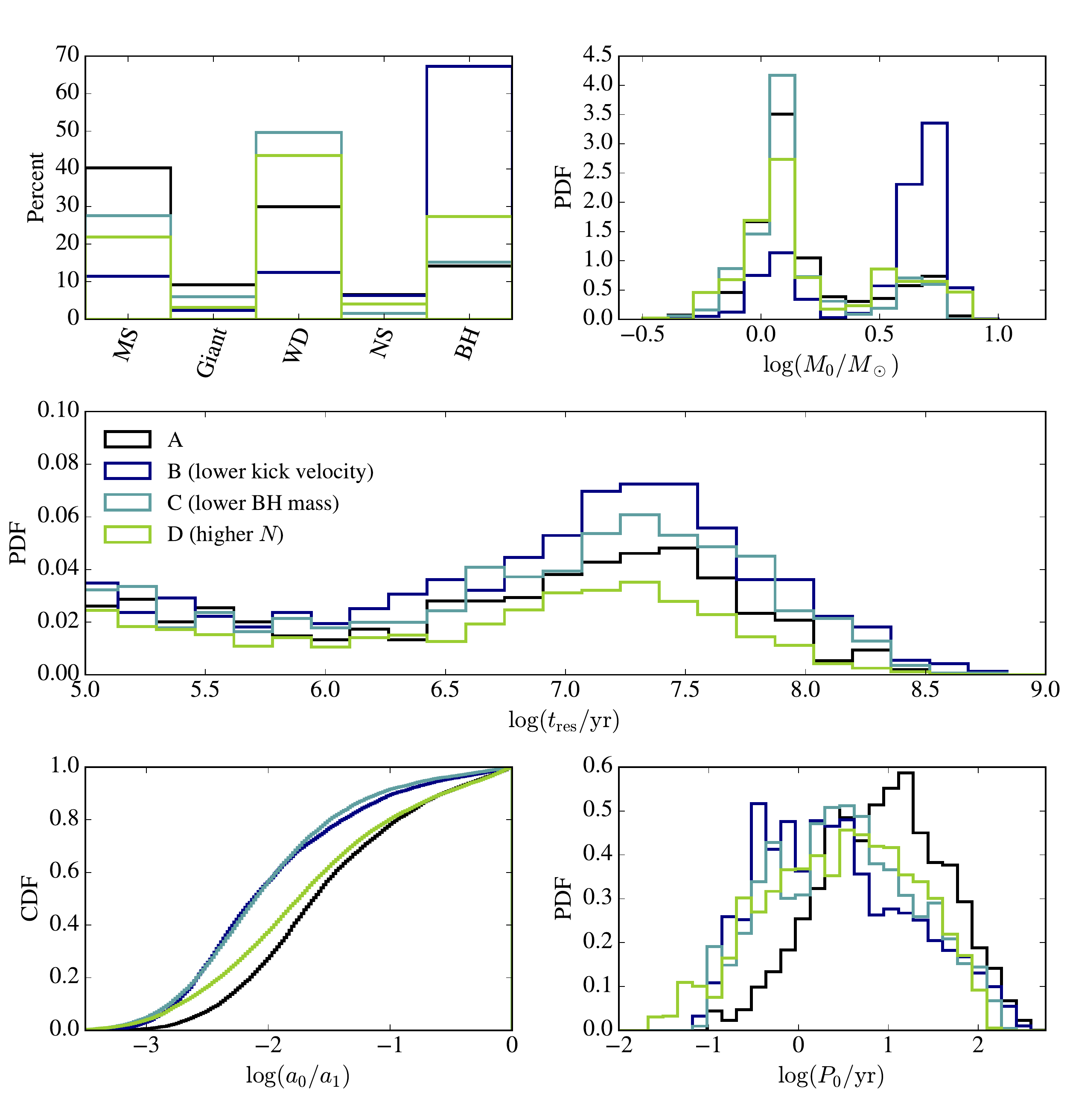}
\caption{In this figure, we show a general comparison of IMBH companion properties with varying simulation initial parameters. We include four groups of simulations in this diagram, A-D, described in Table \ref{table:sims}. Simulation group B, has a lower kick velocity than group A, as a result, more stellar remnant BHs remain in the cluster core. These BHs are more massive than the typical stellar companion, and therefore have longer residence times in companionship with the IMBH because more massive innermost binaries are more difficult to disrupt. As a result of the increased time for the binary to harden, the $a_0/a_1$ distribution extends to smaller separations.  Simulation group C includes a lower IMBH mass of $75M_\odot$ instead of $150M_\odot$, this has only a mild effect on the stellar properties of companions, but does allow companions to segregate to tighter separations in $a_0/a_1$, and thus also shorter median orbital periods. Finally, simulation group D includes twice as many initial particles as the other simulations, $N=200$k. The increased particle number implies shorter residence times as the IMBH's most-bound companion because there are more potential perturbing bodies, and also shorter orbital periods because there are, on average, a larger number of tightly-bound stars.  }
\label{fig:simcomp}
\end{center}
\end{figure}

Figure \ref{fig:simcomp} quantitatively explores how these demographic differences propagate to the other properties of the most-bound companions to the IMBH.  The upper left panel shows companion object demographics, which are also tabulated in Table \ref{table:demo}.  Qualitatively, the demographics remain unchanged in that the companions include stellar remnants like WDs, BHs, and NSs, along with MS and giant stars. As seen in Figures \ref{fig:simcomp_timeseries} and \ref{fig:smademo}, the most dramatic difference arises with simulation group B, and the retention of a higher fraction of BHs in the cluster's central regions. Simulation group D, with its larger total particle number, also exhibits a mild enhancement in the fraction of BH companions. This difference can be traced to the larger total number of BHs retained in these simulations (despite the small retention fraction). Companion demographics shape the companion mass distribution, shown in the upper right panel of Figure \ref{fig:simcomp}. The enhanced BH fraction is reflected in the higher peak at companion masses $\gtrsim 3 M_\odot$ for simulation groups B and D. 

\subsection{Companion Dynamics}

In addition to demographic changes in the typical properties of IMBH companions, there are also dynamical differences we can observe across our simulations. 
We begin by examining how the long-lasting tail of the residence time distribution of IMBH companions is affected by changing cluster properties in the central panel of Figure \ref{fig:simcomp} (the residence time plotted is counted per-partnership as is the grey distribution in Figure \ref{fig:res}). 
This tail of long-lived companion objects is shaped by processes of companion exchange and disruption. As a result, it is influenced by  a combination of cluster, IMBH, and companion demographic properties. 
An increased fraction of BH companions (as in simulation case B) implies more long-lasting partnerships because the IMBH's companion is considerably more immune to exchange  interactions (see Figure \ref{fig:ex}) and to tidal disruption. In fact, we can observe this directly in the termination statistics of Table \ref{table:term}, where we see that exchanges are significantly less frequent in simulation group B (because of the mass sensitivity of the exchange cross-section, Figure \ref{fig:ex}). Tidal disruptions are also nearly a factor of two less common than in case A, with a smaller fraction arising from the most-bound companion. 
With lower IMBH mass, as evaluated in simulation group C,  weaker gravitational focusing of surrounding stars leads to the lower exchange rate tabulated in Table \ref{table:term}. This propagates to more objects with long residence times in the $10^6$ - $10^8$~yr range.
The larger number of stars present in simulation group D, by contrast, implies a larger encounter and exchange rate and shorter typical residence times.

The lower panels of Figure \ref{fig:simcomp} explore differences in companion periods, $P_0$, and hierarchies, $a_0/a_1$. 
Here distinctions arise both in the form of companion demographics and IMBH mass.  All of simulation groups B-D exhibit shorter typical orbital periods than simulation group A. The peak in the orbital period distribution shifts by about 1 dex from $\sim 10$~yr, to $\sim 1$~yr, but the distribution remains broad in all cases. 
Companion hierarchy, as measured by $a_0/a_1$, is also affected both by companion demographics and IMBH mass. The fiducial case, simulation group A, shows companions which are less segregated than groups B-D. In the case of simulation group B, the tighter companion distribution can be traced to the increased BH fraction and correspondingly longer companion residence time (which allows the IMBH-companion pair more time to harden prior to disruption). Simulation group D's tightly-bound tail can be similarly explained by a mildly-increased fraction of BH companions.

The IMBH mass effect on companion hierarchy can also be observed in the lower-left panel of Figure \ref{fig:simcomp}. The median $a_0/a_1$ shifts by a factor of approximately 3 between simulation groups A and C, which have a factor of 2 difference in IMBH mass.  
The origin of this distinction can be understood through a consideration of the dynamics of orbital hardening and exchange for IMBH companions. 
A typical close passage of a perturbing body of mass $M_{\rm p}$ through the IMBH-companion pair carries away an energy of order 
\beq
\Delta E_{\rm p} \approx { G M_{\rm 0} M_{\rm p} \over2  a_0} ,
\eeq
when it passes with orbital pericenter within the semi-major axis of the inner binary \citep{2003ApJ...599.1129Y}. 
As compared to the orbital energy, $E_0 = G \Mbh M_0 / 2 a_0$,  this typical per-encounter change of energy is
\beq
{ \Delta E_{\rm p} \over E_0 }  \approx {  M_{\rm p} \over \Mbh }.
\eeq
So the fractional, per-encounter hardening rate of the IMBH and companion pair scales inversely with the IMBH mass. 
But, the rate of encounters also changes with IMBH mass. The encounter cross section is dictated by gravitational focusing onto the inner binary, and is $\Sigma_{\rm enc} \approx 2 \pi G \Mbh a_0 / v^2$. Associating a typical velocity with the cluster core velocity dispersion, the encounter rate, $\dot N_{\rm enc} \propto n_{\rm c} \sigma_{\rm c} \Sigma_{\rm enc}$, where $n_{\rm c}$ is the core number density,  thus scales as, 
\beq
\dot N_{\rm enc} \propto n_{\rm c} \sigma_{\rm c}^{-1} \Mbh a_{\rm 0},
\eeq
proportional to both $\Mbh$ and $a_{\rm 0}$, as long as the low-angular momentum phase space of encountering orbits is efficiently replenished \citep[e.g.][]{1976MNRAS.176..633F}. As a result, although each encounter with a perturber carries away less energy $\propto \Mbh^{-1}$, the encounters are more common $\propto \Mbh$ and the innermost binary hardens at an identical rate regardless of IMBH mass.

Although the hardening rate remains unchanged with differing IMBH masses, the residence time, as shown in the central panel of Figure \ref{fig:simcomp}, does change. 
In particular, when the encounter rate goes up, so does the exchange rate of IMBH companions. This is because the exchange fraction in our simulations does not appear to depend strongly on IMBH mass, as shown in Figure \ref{fig:ex}, instead it is sensitive primarily to the perturber-companion mass ratio.  
As IMBH masses go up, their companions harden at the same rate, but are exchanged with increasing frequency, restarting a new hardening cycle. 
This result implies, when extrapolated to larger IMBH masses, that only IMBHs of moderate mass ratio to their surrounding stars (perhaps those in the range  $\sim 10^2 -10^4 M_\odot$) will ever exhibit tightly segregated companions.

%
% Discussion
%
\section{Discussion}\label{sec:disc}
In this section we draw on our results concerning the properties of IMBH companions to explore how these objects can occasionally reveal the presence of an otherwise invisible IMBH in the cluster core.

\subsection{IMBH Companions in Old Clusters}

\begin{figure*}[tbp]
\begin{center}
\includegraphics[width=0.99\textwidth]{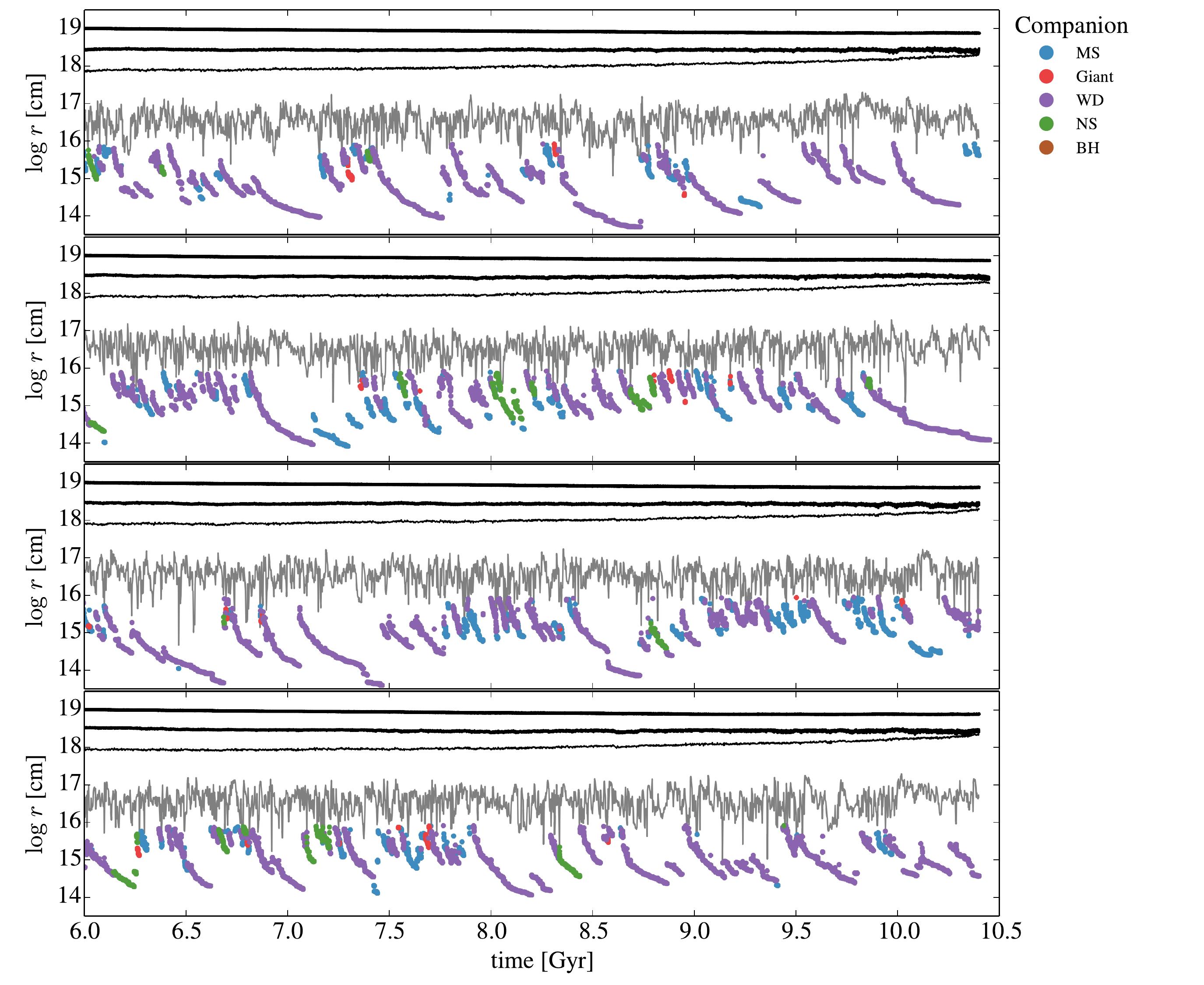}
\caption{Evolution of  representative members the four simulation group D models from 6 Gyr to their tidal dissolution at $\sim10.5$~Gyr. The lines plotted here are a simplified set of those presented in Figure \ref{fig:radii_ks}, from bottom to top in each panel: $a_0$ (color), $a_1$ (grey), $r_{\rm m}$, $r_{\rm c}$, $r_{\rm h}$. This Figure may be compared with the first 6~Gyr of evolution shown in Figures \ref{fig:radii_ks} and \ref{fig:simcomp_timeseries}. IMBH companions in old clusters are qualitatively similar to those at earlier epochs, but contain fewer BHs and a larger fraction of WDs. Sequences for other realizations of a given simulation group are quantitatively different but qualitatively similar to the models reproduced here.  }
\label{fig:radii_late}
\end{center}
\end{figure*}

We have focused so far on IMBH companions across a range of stellar system ages. However, most GCs are  very old stellar systems, with ages $>10$~Gyr \citep[e.g.][]{2009ApJ...694.1498M,2015MNRAS.452.1045F,2015ApJ...808L..35T}. How might the companions to IMBHs in these very old stellar systems differ? To examine this question we look at several evolutionary sequences of the last several Gyr of cluster evolution of the models in simulation group D. These evolutionary tracks are shown in Figure \ref{fig:radii_late}.
We find that by these late epochs interactions with the IMBH (either mergers or ejections) have exhausted the supply of stellar-mass BHs which might fall into partnership with the IMBH \citep{2014MNRAS.444...29L}.
If realistic GCs begin with much larger masses and $N$ than their present-day properties (or our smaller-still simulated systems), they would retain a similar fraction but larger number of BHs to late times \citep{2015ApJ...800....9M}. If these BHs remain in the cluster we would expect them to mass-segregate toward the cluster core and, perhaps,  play a role in shaping the time-dependence of the companion mass function, show in Figure \ref{fig:comp_stellar_prop}.   At 10 Gyr, the turnoff mass is now $\sim 1 M_\odot$, and is therefore less than the mass of typical stellar remnants like WDs and NSs. As a result, the most common companions for the IMBH in our simulations at these times are massive WDs, those with masses $\gtrsim 1 M_\odot$, which are therefore more massive than the MS stars in the cluster core.  NSs similarly form a fraction of late-time companions, because those that were retained in the cluster (but perhaps kicked to the outskirts) have had sufficient time to re-segregate into the cluster core.

\subsection{Dynamical Mechanisms for Repeated Tidal Disruption Flares by IMBHs}
One motivation for performing this study was to access whether close companions to the IMBH are ever driven into multiple-passage tidal interactions. Such multi-passage tidal disruption flaring interactions can be difficult to achieve. For multiple passages of similar strength to be sustained, the star must remain bound to the IMBH with its orbit relatively unchanged (in energy, or more importantly, angular momentum). Further, the object undergoing the passages must expand upon mass loss either through its adiabatic response to mass-loss or heating due to the violent interaction.  
These requirements suggest that orbits leading to multi-passage disruptions must exist in a phase space where the per-orbit scatter due to the remaining cluster stars is small \citep{2013ApJ...777..133M}. For MS stars or WDs in clusters with as few stars as we consider here, the only star in such an orbit is the most-bound companion to the IMBH.  Thus, if this most-bound star enters into a weak tidal encounter with the IMBH the stripping may persist for several passages. 

In our simulations, we find that a reasonable fraction (30-50\%) of tidal disruption events involve the most-bound star (Table \ref{table:term}). Although we track only full disruptions dynamically in our simulations, many of these would, in fact lead to partial tidal disruptions over multiple orbits. Stars with condensed cores relative to their envelopes, in particular evolved stars and those with radiative envelopes are particularly able to retain a bound core despite losing part of their envelope \citep{2012ApJ...757..134M,2013ApJ...767...25G}. We find that some giant stars evolve to the point of mass transfer with the IMBH \citep{2013ApJ...777..133M}. WDs that undergo gravitational wave inspiral are also very likely to disrupt before their orbits circularize, leading to episodic mass transfer and repeated flaring \citep{2014ApJ...794....9M}.  Many of the disrupted most-bound stars reach the tidal radius through a secular interaction with other tightly-bound objects -- opening the possibility of a large number of passages with similar pericenter distance.  In the ensuing repeated tidal stripping episodes the individual flares would be qualitatively similar to other tidal disruption flares albeit with less total fluence. The repetition time between flares would be governed by the orbital period distribution of the most-bound companion. 

This leads us to the case of the extraordinary repeated-flaring transient HLX-1 \citep{2009Natur.460...73F} by a putative IMBH with $\Mbh \sim 10^4 M_\odot$ \citep[e.g.][]{2011ApJ...734..111D}. If this object is powered by the partial tidal stripping of a star in an eccentric orbit the object must have gone through $\gtrsim 10$ passages based on the lightcurve, with orbital period near a year. 
The IMBH mass predicted for HLX-1 is nearly a factor of 100 larger than those we are able to simulate. However, with the scalings of section \ref{sec:diff} as guidance, we can speculate whether these properties appear plausible given our results.

First, having tightly bound stars in orbit around the IMBH appears likely given the IMBH (rather than supermassive) nature of the putative HLX-1 BH. 
Although scalings to a more massive system would modify them, 
 our distributions of orbital periods easily extend to a  orbital periods in the years to tens of years range. 
From a pericenter distance slightly outside the tidal radius, the donor star could have been tidally excited until it started transferring mass to the IMBH \citep{2004ApJ...604L.101H,2006MNRAS.372..467B,2006ApJ...642..427B}.
Secondly, a large fraction of tidal disruption events in our low-density, small-$N$ clusters originate from the most-bound companion to the IMBH (as compared to the supermassive BH in a galactic center context upon which most tidal disruption literature is defined). We observe that the most-bound stars that tend to interact with the IMBH in our simulations arrive in disruptive orbits through secular interaction, rather than single, strong scattering events. Thus, it would not be entirely surprising if a tightly-bound star in the HLX-1 system ended up in a grazing, rather than fully-disruptive, interaction with the IMBH.  In conclusion, a scenario involving a hierarchically isolated companion star being driven to interact with the IMBH is a scenario that qualitatively explains the properties of a repeated flaring source like HLX-1, and extrapolation of our results indicates that such a configuration can be established through dynamical interactions of stars with the IMBH.

\subsection{Revealing IMBHs in GCs}
\subsubsection{Luminous Companions to a Dark IMBH}
The calculations of this paper have demonstrated that IMBHs residing in clusters acquire and retain close companions.  In rare periods in which the IMBH disrupts and accretes its companion, a luminous accretion signature will mark the presence of the IMBH. We discuss this possibility further in the following section, but here we consider the ability of companion objects to reveal the IMBH in periods of relative quiescence.   Among the companion objects, approximately half are luminous stars like MS stars and giants, while the other half are WDs, NSs and smaller BHs.  Can this luminous population of stars serve to reveal the presence of otherwise dark IMBHs in nearby clusters? 

First, the binary nature of the IMBH plus companion pair could potentially reveal the IMBH's presence. The typical companion orbits found in our simulations (semi-major axes of $\sim10^2$ AU and periods of $\sim 10$ yr, shown in Figure \ref{fig:comp_orb_prop}) imply typical orbital velocities of hundreds of km s$^{-1}$. If spectroscopically detected, this orbital motion would strongly point to a massive companion because of the high velocity given combined with relatively long orbital period (compared to a stellar-mass binary).  This approach extends work already being done to explore radial velocities of stars bound to potential IMBHs in cluster cores \citep[e.g.][]{2010ApJ...719L..60N,2013A&A...552A..49L}. But we argue that among the tightly bound stars, it would be reasonable to expect a single very compact companion orbit in a large fraction of clusters hosting a IMBH. 

\begin{figure}[tbp]
\begin{center}
\includegraphics[width=0.49\textwidth]{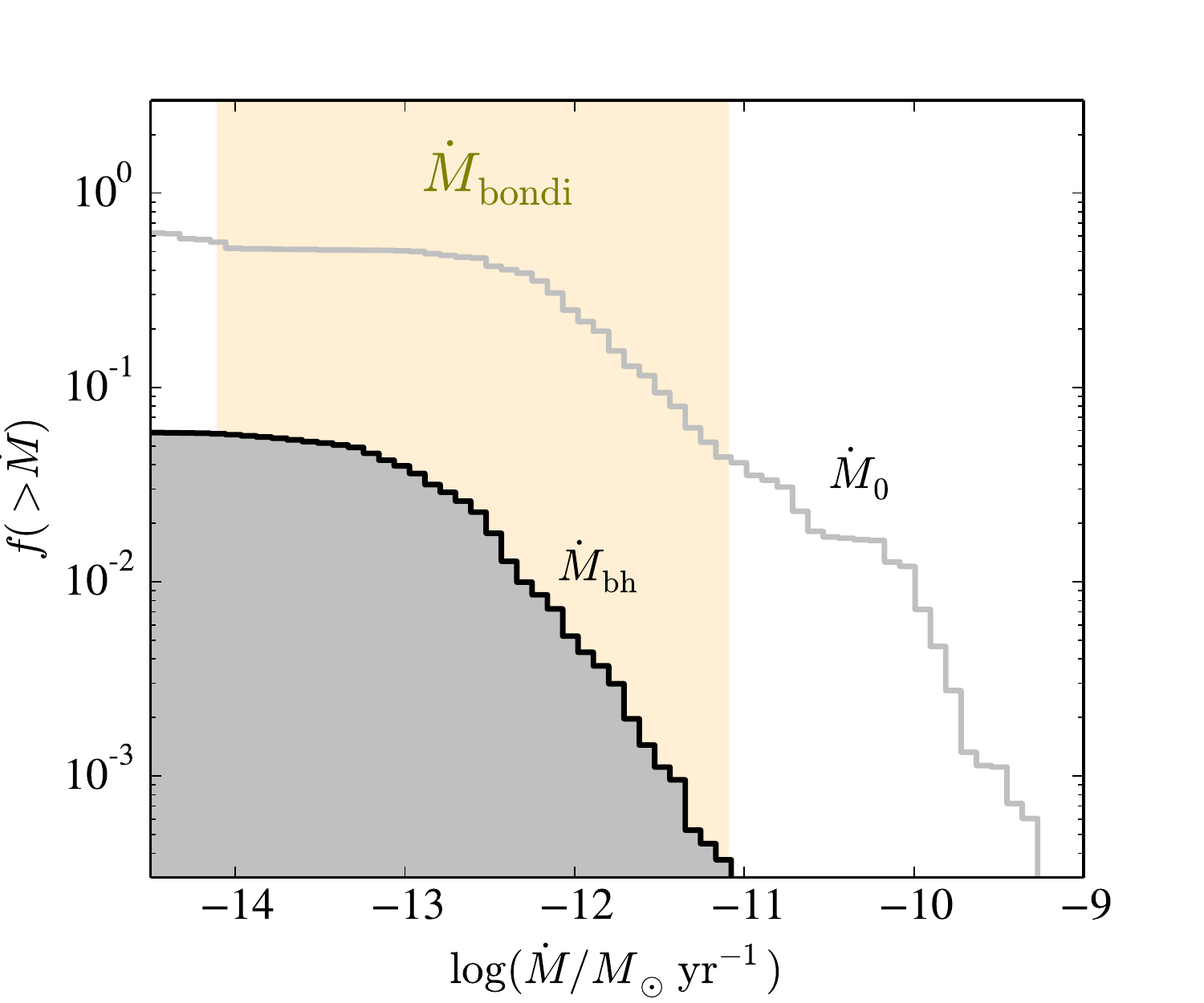}
\caption{The distribution of accretion rates fed by the stellar wind of the companion to the IMBH. The series of lines labeled $\dot M_0$ is the wind mass loss rate of IMBH companions estimated with \citet{1975MSRSL...8..369R}'s formula. The lower series of lines use the wind velocity and orbital properties to estimate the fraction captured by the IMBH. The shaded region denotes the range of possible Bondi accretion rates onto a $150 M_\odot$  IMBH with ambient gas density of $10^{-24}$ g cm$^{-3}$ and sound speed $c_{\rm s} = 10^6 - 10^7$ cm s$^{-1}$ \citep{2012arXiv1206.5002N,2013arXiv1310.8301N}, where $\dot M_{\rm bondi} = 4\pi (G \Mbh)^2 \rho / c_{\rm s}^3$.  Stellar winds are only captured by the IMBH less than 10\% of the time. The typical wind-fed accretion rates during this $\sim5$\% duty cycle are competitive with Bondi accretion from the background of cluster-core stars but only strongly dominant if our more conservative estimate of the gas properties ($c_{\rm s} \sim 10^7$ cm s$^{-1}$ ) applies. Thus, we do not expect companions to IMBHs to impose strong limits on the accretion-fed luminosity of cluster IMBHs.   }
\label{fig:winds}
\end{center}
\end{figure}

Luminous stars also shed substantial stellar winds.  With the star tightly bound to the IMBH we examine whether and how these winds might contribute to fueling accretion activity onto the IMBH.  The relevant scale for comparison here is the Bondi accretion rate at which the IMBH  could accrete gas from the cluster-core interstellar medium, $\dot M_{\rm bondi} = 4\pi (G \Mbh)^2 \rho / c_{\rm s}^3$ \citep{2011ApJ...730..145V}. This interstellar medium is shaped by the interacting stellar winds of all the stars in the cluster core.  Taking typical values of  $10^{-25}$ g cm$^{-3}$ and sound speed $c_{\rm s} = 10^6 - 10^7$ cm s$^{-1}$, we find $\dot M_{\rm bondi} \approx 10^{-11} - 10^{-15} M_\odot$ yr$^{-1}$  \citep[e.g.][]{2012arXiv1206.5002N,2013arXiv1310.8301N}. 

In Figure \ref{fig:winds}, we compare this nominal accretion rate from the background of stars to that fed by the companion.  Here we plot the duty cycle (fraction of time) spent accreting above a given level. The upper series of lines in this figure shows $\dot M_0$, the mass loss rate of the companion star, estimated using \citet{1975MSRSL...8..369R} mass loss formula, $\dot M =  4 \times 10^{-13}  \eta_R (L_0  M_0 / R_0 )/(L_\odot M_\odot/R_\odot)$ with $\eta_R = 0.5$ \citep{2015MNRAS.448..502M}. We then estimate the fraction of this mass lost which is bound to the IMBH using the formulae of \citet[][described in their section 4.1]{2014ApJ...788..116M} The bound fraction depends on the orbital properties through the semi-major axis and eccentricity, which determine the velocity of the wind-shedding star relative to the IMBH. It also depends on the terminal velocity of the winds themselves, which we take to be the star's escape speed, $\sqrt{2 G M_0 / R_0}$.  The slow, massive winds of giant stars, for example, are particularly easy to capture. We plot the portion of companion wind material that is bound to the IMBH as $\dot M_{\rm bh}$ in Figure \ref{fig:winds}.  Most of these captured winds come when the IMBH hosts a giant star companion. These cumulative distributions show that the IMBH captures the winds of its companion at rates comparable to $\dot M_{\rm bondi}$ of order 5\% of the time in our clusters. 

As a result, we can expect that the $\gtrsim 50$\% of stellar IMBH companions should illuminate their host BH through their winds, providing a minimum accretion rate onto the IMBH. 
However, the fact that the IMBH has a companion does not imply that most IMBHs should accrete
 substantially above the Bondi accretion rate from the ambient medium. 
Of course, how these alternative fuel sources transfer into accretion luminosity represents a significant uncertainty. Media captured from the cluster core has low net angular momentum, and likely does not efficiently form a disk \citep[e.g.][]{2008MNRAS.383..458C}, perhaps contributing to a low radiative efficiency. 
There is still more uncertainty in how much of the wind-deposited gas might reach a IMBH in scenarios with realistic feedback \citep{ 2009ApJ...705L.153N,2011ApJ...739....2P,2012ApJ...747....9P,2013ApJ...767..163P} and what fraction of the resultant accretion energy would contribute to radio and X-ray luminosities \citep{2004MNRAS.351.1049M,2005MNRAS.356L..17M,2012ApJ...750L..27S,2015AJ....150..120W}. Given these uncertainties, wind feeding from the companion implies a minimum activity level for the fraction of IMBHs hosting giant star or MS companions.

\subsubsection{Electromagnetic and Gravitational Transients}
Stellar tidal disruption events lead to a stream of debris from the disrupted star accreting onto the IMBH. The ensuing accretion flare may be extremely luminous \citep{2009ApJ...697L..77R}.  For the relatively low-mass clusters with low-mass IMBHs we have considered through our direct $N$-body integrations, the rate of tidal disruptions is relatively low, $\sim 10^{-8} $ yr$^{-1}$. More massive IMBHs and denser clusters both imply higher event rates, with $\dot N_{\rm tde} \propto \Mbh^{4/3} n_{\rm c}$ \citep{1988Natur.333..523R}. The events themselves, however, are likely short, with the flare characteristic timescales of order weeks to months \citep[e.g.][]{2009ApJ...697L..77R,2012ApJ...757..134M,2013ApJ...767...25G} and observable signatures lasting at most for years or tens of years through nebular emission \citep{2011ApJ...726...34C,2012MNRAS.424.1268C}.  The implied duty cycle of these flaring episodes is thus extremely small and the fraction of cluster IMBHs expected to be in a flaring state is $\sim 10^{-8}$. Tracing this number of clusters is not inconceivable, optical surveys for tidal disruption flares routinely cover $\sim10^5$ galaxies \citep[e.g.][]{2011ApJ...741...73V}, each of which might host $10^2-10^3$ globular clusters \citep{2006ARA&A..44..193B}.  But this scaling argument is hampered by the fact that the Eddington-limited luminosity of low mass IMBHs is much lower than their supermassive counterparts. Thus, this argument only suggests a reasonable fraction of detectable events if the occupation fraction of IMBHs in local-universe clusters is of order unity. 
 
Relativistically beamed jet emission from tidal disruption events may be visible in some cases, and for low mass BHs, the implied luminosities are large enough to suggest that these events could be observed to cosmological distances \citep{2011Sci...333..203B,2011Natur.476..425Z,2012ApJ...748...36B,2012ApJ...760..103D,2013ApJ...767..152Z}. Because of their high apparent luminosity these high-energy transients, especially those arising from rare but luminous WD tidal disruptions, could dominate the observable population of tidal disruption flaring events arising from IMBHs \citep{2011ApJ...743..134K,2014ApJ...794....9M,2015arXiv150802399M}.  With a population of these transients we might eventually hope to understand the cosmological demographics and temporal evolution of IMBHs in clusters. But given their extreme rarity, these events do little to probe any potential local-group population of IMBHs in GCs. 
 
Gravitational wave inspirals do offer a promising new probe of the presence or absence of IMBHs in nearby clusters \citep[e.g.][]{2002MNRAS.330..232C,2004ApJ...616..221G,2005MNRAS.363L..56H,2007CQGra..24R.113A,2007A&A...476..121I,2008MNRAS.391..718S,2008ApJ...681.1431M,2009CQGra..26i4036M,2013A&A...557A.135K}. With laser interferometer experiments like advanced LIGO entering a phase of enhanced sensitivity, the possible detection of gravitational radiation from a merging binary appears promising. How might gravitational wave inspirals of compact objects into cluster IMBHs compare? Because of their higher masses, IMRI events enter the LIGO band only near the end of the inspiral \citep{2002MNRAS.330..232C}. For example, the gravitational wave frequency at the innermost stable circular orbit of a $100 M_\odot$ IMBH is $\sim 40$ Hz, with an amplitude that might allow its observation to $z\sim 0.1$ \citep{2008ApJ...681.1431M}. Similarly, \citet{2013A&A...557A.135K} show that a LISA-like space-based detector could capture lower-frequency emission from IMRIs at times several years prior merger out to $z \sim 0.7$.

The event rate of IMRIs compared to standard binary mergers is much more uncertain.
In our simulations, we find similar IMRI rates per GC, $\sim 1$ Gyr$^{-1}$, as estimated previously by \citet{2008ApJ...681.1431M}. 
For context,  we can assume a galaxy density of $10^7$ Gpc$^{-3}$ and $10^2$ globular clusters per galaxy \citep{2015MNRAS.452..575S,2006ARA&A..44..193B}.  If we conservatively take the inspiral rate from our low-mass clusters as representative, then an inspiral rate of order $1$ yr$^{-1}$  Gpc$^{-3}$ is expected if every cluster hosts a IMBH.  
This is a factor of $\sim 100$ lower than the double NS inspiral rate estimated recently \citep[e.g.][]{2013ApJ...779...72D,2015ApJ...806..263D}, but the large chirp masses of these IMBH binaries make their inspirals  observable in the larger volumes mentioned above that might imply similar detection rates of both classes of source \citep{2002MNRAS.330..232C,2008ApJ...681.1431M} These constraints are compelling enough to offer hope that the upcoming era of gravitational wave detections will either offer guiding detections or constraining non-detections to pin down the IMBH occupation fraction in low-redshift clusters.

\subsection{Extensions and Future Work}
We have explored the question of close companion objects to IMBHs in dense cluster environments using direct $N$-body numerical simulations in this paper. 
This chosen approach carries the advantage of direct integration of each particle's equation of motion, ensuring accurate dynamics. However, it  necessitates some approximations due to the computational expense of each individual simulation. In particular, in our analysis we consider clusters with initial mass of $\sim 5 \times 10^4  M_\odot$ or $10^5 M_\odot$, by the time these clusters evolve, they are less massive (and thus less dense) than the the typical Milky Way GC.  We similarly include IMBHs of relatively low initial mass, $75$ or $150M_\odot$. More massive clusters might harbor more-massive still IMBHs, changing the IMBH to star mass ratio.  While these constraints are shared by every direct $N$-body simulation method, they necessitate a consideration of how the stellar dynamics will scale and extrapolate to more massive systems found in Nature. We have qualitatively explored the response and dependence of our results to dependencies in cluster mass and density and IMBH mass in Section \ref{sec:diff}.  To extend our simulated dynamics into the regimes that found in more massive IMBH environments,  an $N$-body algorithm particularly well suited to study more extreme mass ratios, like that recently published by \citet{2015MNRAS.452.2337K}, may offer a promising numerical approach. 

We have included no primordial binaries in the simulations presented in this work. This simplification is not realistic; binary fractions are thought to be of order a few percent in typical GCs \citep{2003gmbp.book.....H}. In future work, we intend to include a realistic binary fraction and study the role these primordial binaries may play in interacting with the IMBH and its close companions \citep{2005ApJ...626..849P}. Before addressing this question with numerical simulations, we can make a few analytic estimates of their potential importance. In a cluster environment, only hard binaries survive the interactive cluster core environment \citep{1975MNRAS.173..729H}. To order of magnitude, these binaries have separation $G M_{\rm bin} / \sigma_{\rm c}^2$. If such a binary passes close to the IMBH and is split by the \citet{1988Natur.331..687H} mechanism, the one star can remain bound to the IMBH while the other is ejected. The bound star has a minimum typical semi-major axis of 
\beq
a_{\rm min} \sim \left( \Mbh \over M_{\rm bin} \right)^{2/3} a_{\rm bin},
\eeq
plugging in a few typical numbers for our simulation clusters, this might indicate $a_{\rm min} \sim 3 \times 10^{15}$~cm, for marginally-hard source binaries.  This is near the peak of the semi-major axis distribution of IMBHs companions in our case A simulations (as seen, for example, in Figure \ref{fig:comp_orb_prop}). Thus, in some cases, these split binaries would leave behind stars with similar semi-major axis to the most-bound IMBH companion. Additional strong interactions between the newly-bound star and the companion would ensue with the possibility of an exchange of hierarchy. 

\citet{2007MNRAS.374..857T} run direct N-body simulations of star
clusters with primordial binaries and a central IMBH under idealized
conditions ($N=16384$ equal-mass particles) and find that the IMBH
presence increases the binary disruption rate in qualitative agreement
with the theoretical predictions by \citet{2005ApJ...626..849P}. However, the
large majority of the binaries are split by interactions with other
stars in the dense central stellar cusp surrounding the IMBH, rather
than by the IMBH tidal field \citep[][Figure 6]{2007MNRAS.374..857T}. This hints that a
majority of binaries in relatively low angular momentum orbits may be
split before reaching the IMBH, and that those that survive as bound pairs
and experience a close encounter with the IMBH are likely to be
preferentially harder binaries. The earlier idealized experiments of
IMBH-binary interactions also highlight the importance of resolving
the dynamics of the sphere of influence exactly through direct N-body
simulations. We plan to explore these effects more fully with the aid
of future simulations.

\section{Summary \& Conclusion}\label{sec:conclusion}

IMBHs residing in clusters acquire close companions through strong dynamical interactions with their surroundings. We use full $N$-body calculations including stellar evolution to 
present a systematic study of the demographics, dynamics, and observable properties of these IMBH companions. 
This work confirms suggestions that IMBHs in dense clusters should have close companions for the majority of their evolution   \citep{2006ApJ...642..427B,2013MNRAS.429.2298M,2014ApJ...794....7M}. We find that $\sim90$\% of the time, companion objects have semi major axes less than a third that of the next-most bound object, $a_{0} < 1/3 a_1$.  The hierarchy can be substantially more extreme, with a few percent of companion configurations having $a_0 \sim 10^{-3} a_1$ (Figure \ref{fig:comp_orb_prop}). The most-bound object demographics are broad, including a wide variety of stars and compact objects. 

We show that most-bound partners to the IMBH go through cycles of orbital hardening followed by destruction, illustrated in Figure \ref{fig:radii_ks}. The typically-observed pairings are long-lived, with residence times of $10^6-10^8$~yr, shown in Figure \ref{fig:res}. These pairings are terminated through exchange or close interaction with the IMBH (gravitational wave inspiral or tidal disruption). We diagram the possible channels through which a partnership can be terminated in Figure \ref{fig:diagram}, and show the destructive close encounters with the IMBH on the cluster evolution timeseries in Figure \ref{fig:radii_dis}. 

In Section \ref{sec:diff}, we varied several key cluster parameters to consider the sensitivity of IMBH companions to cluster and IMBH parameters. We considered clusters with a higher stellar remnant retention fraction, smaller IMBH mass, and larger particle number as compared to our fiducial parameter set. These comparisons show varying compositions of companion stellar type demographics (Figure \ref{fig:simcomp}). In particular, the retention of more NSs and BHs leads to a substantially larger fraction of BH companions as these objects mass-segregate to interaction with the IMBH \citep[e.g.][]{2014MNRAS.444...29L}. These demographic differences propagate to minor changes in typical companion orbital semi-major axis (or period) as shown in Figure \ref{fig:smademo}, where we break up the semi-major axis distribution of companions by stellar type. We show that companion hierarchy  (as defined by $a_0/a_1$) depends on IMBH mass. Lower-mass IMBHs show companions that are more highly segregated from the remaining stellar system than do more massive IMBHs. This suggests that the capture of hierarchically segregated companions is a unique property of the IMBH mass range. 

The segregation of close companions to IMBHs, as a property unique to BHs in the intermediate mass range,  has dramatic implications for the production of repeated flaring episodes in tidal interactions between stars and IMBHs, as seen in the remarkable object HLX-1 \citep{2009Natur.460...73F}. A combination of stellar dynamical segregation, secular perturbation from less bound stars, and tidal interaction with the IMBH \citep[e.g.][]{2004ApJ...604L.101H,2006MNRAS.372..467B,2006ApJ...642..427B} could place a star in an orbit similar to the 1~yr period observed for HLX-1.  Continued interactions with the cluster background could then account for the perturbation of this orbit into a disruptive configuration. 

In closing, we note that we may be able to leverage detailed knowledge of IMBH  companions to reveal or constrain the still-uncertain presence of IMBHs in dense clusters including  GCs, young clusters, and compact galactic nuclear clusters.  
IMBHs themselves are notoriously tricky to uncover, but our analysis suggests that they may reveal themselves several transient channels. Gravitational waves generated by IMRI events should carry large enough amplitudes to be detectable to moderate redshifts. Our results indicate that these events occur, even in the modeled low-mass cluster, at rates similar to analytic predictions \citep{2008ApJ...681.1431M}, offering substantial promise for their detection by either ground or space based laser interferometers. 
Electromagnetic emission might arise through steady-state accretion from companion winds.  Our calculations, shown in Figure \ref{fig:winds}, suggest that this may exceed the IMBHs Bondi accretion rate from the cluster medium only a small fraction, $\sim5$\%, of the time, but it none-the-less suggests that a lower limit accretion rate exists for IMBHs in dense clusters. 
Finally, we discuss extremely luminous transient accretion occurring in rare episodes of tidal disruption of stars by the IMBH. Especially if the IMBH launches jets in these events, these transients could trace the presence of  IMBHs in clusters to cosmological distances.

\begin{acknowledgements}
We are grateful to Sverre Aarseth for many guiding conversations and advice on the inclusion of IMBHs and new diagnostics in star cluster simulations with NBODY6; we are lucky to have this fantastic code resource at our disposal. We thank the participants of the Stellar $N$-body Dynamics conference in Sexten, Italy, for helpful questions and discussion. 
We are grateful to Melvyn Davies, Shawfeng Dong, Jacqueline Goldstein, James Guillochon, Phillip Macias, Michela Mapelli, Cole Miller, Jill Naiman, Martin Rees, Johan Samsing, Mario Spera, Rainer Spurzem, Luca Zampieri and Brunetto Ziosi for advice and helpful conversations. 
We thank the anonymous referee for constructive and useful feedback. 
This research made use of \texttt{Astropy}, a community-developed core Python package for Astronomy \citep{2013A&A...558A..33A}.
The simulations for this research were carried out on the UCSC
supercomputer Hyades, which is supported by National Science Foundation (award number AST-1229745) and UCSC. 
MM is grateful for the support of the Chancellor's Fellowship at UCSC. ER-R acknowledges financial support from the Packard Foundation, Radcliffe Institute for Advanced Study and  NASA ATP grant NNX14AH37G.

\end{acknowledgements}

\bibliographystyle{apj}
%\bibliography{bhcomp}

\end{document}